\documentclass[english,aip,pop,reprint]{revtex4-1}
\usepackage{lmodern}
\pdfoutput=1

\usepackage[T1]{fontenc}
\usepackage[utf8]{inputenc}
\usepackage{color}
\usepackage{babel}
\usepackage{units}
\usepackage{mathrsfs}
\usepackage{mathtools}
\usepackage{amsmath}
\usepackage{amssymb}
\usepackage[unicode=true,
 bookmarks=true,bookmarksnumbered=false,bookmarksopen=false,
 breaklinks=false,pdfborder={0 0 0},backref=false,colorlinks=true]
 {hyperref}
\hypersetup{
 pdfauthor={Rudi Gaelzer},
 linkcolor=blue, citecolor=blue}

\makeatletter
\allowdisplaybreaks

\makeatother

\begin{document}
\global\long\def\pFq#1#2{\prescript{\vphantom{#1}}{#1}{F}_{#2}^{\vphantom{#2}}}

\title{The general dielectric tensor for bi-kappa magnetized plasmas}

\author{R. Gaelzer}

\email{rudi.gaelzer@ufrgs.br}

\author{L. F. Ziebell}

\email{luiz.ziebell@ufrgs.br}

\author{A. R. Meneses}

\email{anemeneses@gmail.com}

\affiliation{Instituto de Física, UFRGS, 91501-970 Porto Alegre, RS, Brazil}

\begin{abstract}
In this paper we derive the dielectric tensor for a plasma containing
particles described by an anisotropic superthermal (bi-kappa) velocity
distribution function. The tensor components are written in terms
of the two-variables kappa plasma special functions, recently defined
by Gaelzer and Ziebell {[}Phys. Plasmas \textbf{23}, 022110 (2016){]}\@.
We also obtain various new mathematical properties for these functions,
which are useful for the analytical treatment, numerical implementation
and evaluation of the functions and, consequently, of the dielectric
tensor. The formalism developed here and in the previous paper provides
a mathematical framework for the study of electromagnetic waves propagating
at arbitrary angles and polarizations in a superthermal plasma.
\end{abstract}
\maketitle

\section{Introduction}

During the last years, a substantial portion of the space physics
community has been interested in plasma environments which are not
in a state of thermal equilibrium, but are instead in a turbulent
state. Several of such environments can be found in a nonthermal (quasi-)
stationary state. When the velocity distribution functions (VDFs)
of the particles that comprise these turbulent plasmas are measured,
they often display a high-energy tail that is better fitted by a power-law
function of the particle's velocity, instead of the Gaussian profile
found in plasmas at the thermodynamic equilibrium. 

Among all possible velocity distributions with a power-law tail, the
actual VDF that has been marked with a widespread application in space
plasmas is the Lorentzian, or kappa, distribution (or a combination
of kappas), and the number of published papers that employ the kappa
velocity distribution function ($\kappa$VDF) has been growing by
a measurable exponential rate.\cite{LivadiotisMcComas11/11} However,
the interest on the kappa distribution is justified not only as a
better curve-fitting function. A kappa function is also the velocity
probability distribution that results from the maximization of the
nonadditive Tsallis entropy postulate. Hence, the $\kappa$VDF is
also the distribution of velocities predicted by Tsallis's entropic
principle for the nonthermal stationary state of a statistical system
characterized by low collision rates, long-range interactions and
strong correlations among the particles. For detailed discussions
of the importance of kappa distributions for space plasmas and the
connection with nonequilibrium statistical mechanics, the Reader is
referred to Refs. \onlinecite{LivadiotisMcComas09/11,LivadiotisMcComas11/11,LivadiotisMccomas13/05,Livadiotis15/03}.
See also the Introductions of Refs. \onlinecite{GaelzerZiebell14/12,GaelzerZiebell16/02}
for complementary discussions and references to other formulations
for the $\kappa$VDF.

One of the important problems related to space plasmas in which the
kappa distribution has been increasingly applied concerns the excitation
of temperature-anisotropy-driven instabilities that propagate in electromagnetic
or electrostatic modes in a warm plasma. These instabilities (among
others) play an important role on the nonlinear evolution and the
steady-state of the measured VDFs. They can also lead to particle
energization and acceleration and are probably related to some of
the fundamental issues in space and astrophysical systems, such as
the problem of the heating of the solar corona. Rather than giving
here a long list of references, we suggest that the Reader consults
the cited literature in our previous works.\cite{GaelzerZiebell14/12,GaelzerZiebell16/02}

In the present work, we continue the development of a mathematical
formulation destined to the study of electromagnetic/electrostatic
waves (and their instabilities) propagating at arbitrary angles in
a warm magnetized plasma, in which the particles are described by
asymmetric superthermal, or bi-kappa, VDFs. The formulation presented
here employs the linear kinetic theory of plasmas and is an extension
and generalization of the treatment developed in Refs. \onlinecite{GaelzerZiebell14/12,GaelzerZiebell16/02}.

The structure of this paper is as follows. In Section \ref{sec:GZM16:Diel_Tensor}
we derive the dielectric tensor for a bi-kappa plasma. The tensor
components are written in terms of the kappa plasma special functions
introduced and studied in our previous works. Section \ref{sec:GZM16:Kappa_plasma_functions}
contains several new developments and properties of the kappa plasma
functions, destined to provide the necessary framework for the evaluation
of the functions and the dielectric tensor. After the conclusions
in Section \ref{sec:Conclusions}, we have also included Appendices
\ref{sec:GZM16:SuscTensor-derivation}, where details about the derivation
of the dielectric tensor are given, and \ref{sec:GZM16:G-functions},
where additional properties of relevant special functions are derived.

\section{The dielectric tensor\label{sec:GZM16:Diel_Tensor}}

The dielectric tensor for a bi-kappa superthermal plasma will be obtained
with the use of the velocity distribution function given by 
\begin{equation}
f_{s}^{\left(\alpha\right)}\left(v_{\parallel},v_{\perp}\right)=A_{s}^{(\sigma_{s})}\left(1+\frac{v_{\parallel}^{2}}{\kappa_{s}w_{\parallel s}^{2}}+\frac{v_{\perp}^{2}}{\kappa_{s}w_{\perp s}^{2}}\right)^{-\sigma_{s}},\label{eq:KAP7:f_s}
\end{equation}
which is valid for $\sigma_{s}>\nicefrac{3}{2}$ and where $\sigma_{s}=\kappa_{s}+\alpha_{s}$
and 
\[
A_{s}^{(\sigma_{s})}=\frac{1}{\pi^{3/2}w_{\parallel s}w_{\perp s}^{2}}\frac{\kappa_{s}^{-3/2}\Gamma\left(\sigma_{s}\right)}{\Gamma\left(\sigma_{s}-\nicefrac{3}{2}\right)}
\]
is the normalization constant. The quantities $w_{\parallel s}$ and
$w_{\perp s}$ are respectively proportional to the parallel and perpendicular
thermal speeds, but they can be a function of the $\kappa$ parameter
as well. Finally, $\Gamma\left(z\right)$ is the gamma function. 

The VDF (\ref{eq:KAP7:f_s}) is the anisotropic generalization of
the isotropic $\left(w_{\parallel}=w_{\perp}=w\right)$ distribution
adopted by Refs. \onlinecite{GaelzerZiebell14/12,GaelzerZiebell16/02}\@.
In these works, it was shown how adequate choices of the parameters
$\alpha$ and $w$ reproduce and formally unify seemingly different
(kappa) velocity distribution functions employed in the literature.
Now, in the anisotropic case, if we set $\alpha=1$ and 
\[
w_{\parallel,\perp}^{2}=\theta_{\parallel,\perp}^{2}=\left(1-\frac{3}{2\kappa}\right)\left(\frac{2T_{\parallel,\perp}}{m}\right),
\]
the function (\ref{eq:KAP7:f_s}) reduces to the ``bi-Lorentzian''
distribution introduced by Summers \& Thorne\cite{SummersThorne91/08}
(see Table I\@. See also Eqs. 12a,b of Ref. \onlinecite{Livadiotis15/03})\@.
This distribution will be named here the ST91 model and in all expressions
obtained below one can simply drop the parameter $\alpha$, should
the ST91 model be chosen from the start.

However, the parameter $\alpha$ can also be useful when the function
$f_{s}^{\left(\alpha\right)}\bigl(v_{\parallel},v_{\perp}\bigr)$
describes (isotropic) one-particle distribution functions with an
arbitrary number of degrees of freedom. If (\ref{eq:KAP7:f_s}) describes
the probability distribution function of a particle with $f$ degrees
of freedom, one can set $\kappa=\kappa_{0}$, where $\kappa_{0}$
is the \emph{invariant} kappa parameter introduced by Livadiotis \&
McComas\cite{LivadiotisMcComas11/11}, $\alpha=1+f/2$, $w^{2}=\theta^{2}=2T/m$,
$v^{2}=\sum_{i=1}^{f}v_{i}^{2}$, and the normalization constant is
$A^{(f)}=\Gamma\left(\kappa_{0}+1+f/2\right)\left(\pi\kappa_{0}\theta^{2}\right)^{-f/2}/\Gamma\left(\kappa_{0}+1\right)$,
thereby obtaining Eq. (22c) of Ref. \onlinecite{LivadiotisMcComas11/11}\@.

Particular forms of the bi-kappa VDF (\ref{eq:KAP7:f_s}) or its bi-Maxwellian
limiting case (when $\kappa_{s}\to\infty$) have been frequently employed
in the literature in order to study temperature-anisotropy-driven
instabilities that amplify parallel- or oblique-propagating eigenmodes
in a magnetized plasma. Of particular importance for the present work
are the effects of finite particle gyroradius (or Larmor radius) on
the dispersion and amplification/damping of oblique-propagating modes.
For instance, Yoon \emph{et al}.\cite{YoonWuAssis93/07} discovered
the oblique Firehose instability, which is a non-propagating instability
excited in a high-beta bi-Maxwellian plasma when the ions display
temperature anisotropy (with $T_{\parallel i}>T_{\perp i}$) and which
is continuously connected to the left-handed branch of the Alfvén
waves when the ion gyroradius tends to zero. The same instability
was later rediscovered by Hellinger \& Matsumoto.\cite{HellingerMatsumoto00/05}

Other studies subsequently considered the excitation of low-frequency
instabilities at arbitrary angles in bi-Maxwellian plasmas for other
situations, such as low-beta plasmas,\cite{Bashir+10/10} or with
additional free energy sources such as electronic temperature anisotropy,\cite{ChenWu10/06}
field-aligned currents,\cite{Chen+13/06} loss-cones,\cite{Shukla+08/02}
and density inhomogeneities\cite{Naim+14/03} (see also reviews by
Refs. \onlinecite{BashirMurtaza12/12,KleinHowes15/03,Gary15/04}).

In comparison, similar studies employing anisotropic superthermal
distributions are rare. Summers \emph{et al.}\cite{Summers+94/06}
obtained the first expressions for the general dielectric tensor of
a bi-kappa (bi-Lorentzian) plasma. However, their final expressions
are not written in a closed form, i.e., for each component of the
tensor, there remains a final integral along $v_{\perp}$ (the perpendicular
component of the particle's velocity) that should be numerically evaluated.
A similar approach was later adopted by Basu,\cite{Basu09/05} Liu
\emph{et al}\@.\cite{Liu+14/03} and Astfalk \emph{et al}\@.\cite{Astfalk+15/09}

In order to circumvent the mathematical difficulties involved in the
integration along $v_{\perp}$, Cattaert \emph{et al}.\cite{Cattaert+07/08}
derived the dielectric tensor and considered some simple cases of
oblique waves propagating in a kappa-Maxwellian plasma. More recently,
Sugiyama \emph{et al.}\cite{Sugiyama+15/10} employed the same VDF
in a first systematic study of the propagation of electromagnetic
ion-cyclotron waves in the Earth's magnetosphere.

Closed-form expressions for the components of the dielectric tensor
of a superthermal plasma were for the first time obtained by Gaelzer
\& Ziebell,\cite{GaelzerZiebell14/12,GaelzerZiebell16/02} still for
the particular case of isotropic $\left(w_{\parallel s}=w_{\perp s}\right)$
distributions. Here, we will obtain the dielectric tensor for the
bi-kappa VDF given by (\ref{eq:KAP7:f_s}).

The general form of the dielectric tensor can be written as\cite{GaelzerZiebell16/02}\begin{subequations}\label{eq:DielTensor-general}
\begin{align}
\varepsilon_{ij}\left(\boldsymbol{k},\omega\right)= & \delta_{ij}+\sum_{s}\chi_{ij}^{\left(s\right)}\left(\boldsymbol{k},\omega\right),\label{eq:KAP7:DielTensor-1}\\
\chi_{ij}^{\left(s\right)}\left(\boldsymbol{k},\omega\right)= & \frac{\omega_{ps}^{2}}{\omega^{2}}\left[\sum_{n\to-\infty}^{\infty}\int d^{3}v\frac{v_{\perp}\left(\Xi_{ns}\right)_{i}\left(\Xi_{ns}^{*}\right)_{j}\mathcal{L}f_{s}}{\omega-n\Omega_{s}-k_{\parallel}v_{\parallel}}\right.\nonumber \\
 & \left.+\vphantom{\sum_{n\to-\infty}^{\infty}}\delta_{iz}\delta_{jz}\int d^{3}v\frac{v_{\parallel}}{v_{\perp}}Lf_{s}\right]\label{eq:SuscTensor}
\end{align}
\end{subequations}where $\chi_{ij}^{\left(s\right)}$ is the susceptibility
tensor associated with particle species $s$, the set $\left\{ i,j\right\} =\left\{ x,y,z\right\} $
identifies the Cartesian (in the $E^{3}$ space) components of the
tensors, with $\left\{ \hat{\boldsymbol{x}},\hat{\boldsymbol{y}},\hat{\boldsymbol{z}}\right\} $
being the basis in $E^{3}$, $\boldsymbol{\Xi}_{ns}=n\varrho_{s}^{-1}J_{n}\left(\varrho_{s}\right)\hat{\boldsymbol{x}}-iJ_{n}'\left(\varrho_{s}\right)\hat{\boldsymbol{y}}+\left(v_{\parallel}/v_{\perp}\right)J_{n}\left(\varrho_{s}\right)\hat{\boldsymbol{z}}$,
where $J_{n}\left(z\right)$ is the Bessel function of the first kind,\cite{OlverMaximon-NIST10,NIST10}
$\varrho_{s}=k_{\perp}v_{\perp}/\Omega_{s}$, $Lf_{s}=v_{\perp}\partial f_{s}/\partial v_{\parallel}-v_{\parallel}\partial f_{s}/\partial v_{\perp}$,
$\mathcal{L}f_{s}=\omega\partial f_{s}/\partial v_{\perp}+k_{\parallel}Lf_{s}$\@.
Also, $\omega_{ps}^{2}=4\pi n_{s}q_{s}^{2}/m_{s}$ and $\Omega_{s}=q_{s}B_{0}/m_{s}c$
are the plasma and cyclotron frequencies of species $s$, respectively,
$\omega$ and $\boldsymbol{k}=k_{\perp}\hat{\boldsymbol{x}}+k_{\parallel}\hat{\boldsymbol{z}}$
are the wave frequency and wavenumber, $\boldsymbol{B}_{0}=B_{0}\hat{\boldsymbol{z}}$
$\left(B_{0}>0\right)$ is the ambient magnetic induction field and
the symbols $\parallel\left(\perp\right)$ denote the usual parallel
(perpendicular) components of vectors/tensors, respective to $\boldsymbol{B}_{0}$\@. 

Inserting the function (\ref{eq:KAP7:f_s}) into (\ref{eq:SuscTensor})
we obtain the desired susceptibility tensor for a bi-kappa plasma.
More details on the derivation of the components of $\chi_{ij}$ are
given in Appendix \ref{sec:GZM16:SuscTensor-derivation}\@. Here
we will presently show the final, closed-form expressions, given by\begin{subequations}\label{eq:SuscTensor_Bi-Kappa}
\begin{align}
\chi_{xx}^{\left(s\right)}= & \frac{\omega_{ps}^{2}}{\omega^{2}}\sum_{n\to-\infty}^{\infty}\frac{n^{2}}{\mu_{s}}\left[\vphantom{\frac{1}{2}}\xi_{0s}\mathcal{Z}_{n,\kappa_{s}}^{\left(\alpha_{s},2\right)}\left(\mu_{s},\xi_{ns}\right)\right.\nonumber \\
 & \left.+\frac{1}{2}A_{s}\partial_{\xi_{ns}}\mathcal{Z}_{n,\kappa_{s}}^{\left(\alpha_{s},1\right)}\left(\mu_{s},\xi_{ns}\right)\right]\label{eq:ST-BK-1}\\
\chi_{xy}^{\left(s\right)}= & \,i\frac{\omega_{ps}^{2}}{\omega^{2}}\sum_{n\to-\infty}^{\infty}n\left[\vphantom{\frac{1}{2}}\xi_{0s}\partial_{\mu_{s}}\mathcal{Z}_{n,\kappa_{s}}^{\left(\alpha_{s},2\right)}\left(\mu_{s},\xi_{ns}\right)\right.\nonumber \\
 & \left.+\frac{1}{2}A_{s}\partial_{\mu_{s},\xi_{ns}}^{2}\mathcal{Z}_{n,\kappa_{s}}^{\left(\alpha_{s},1\right)}\left(\mu_{s},\xi_{ns}\right)\right]\\
\chi_{xz}^{\left(s\right)}= & -\frac{\omega_{ps}^{2}}{\omega^{2}}\frac{w_{\parallel s}}{w_{\perp s}}\sum_{n\to-\infty}^{\infty}\frac{n\Omega_{s}}{k_{\perp}w_{\perp s}}\left(\xi_{0s}-A_{s}\xi_{ns}\right)\nonumber \\
 & \times\partial_{\xi_{ns}}\mathcal{Z}_{n,\kappa_{s}}^{\left(\alpha_{s},1\right)}\left(\mu_{s},\xi_{ns}\right)\\
\chi_{yy}^{\left(s\right)}= & \frac{\omega_{ps}^{2}}{\omega^{2}}\sum_{n\to-\infty}^{\infty}\left[\xi_{0s}\mathcal{W}_{n,\kappa_{s}}^{\left(\alpha_{s},2\right)}\left(\mu_{s},\xi_{ns}\right)\right.\nonumber \\
 & \left.+\frac{1}{2}A_{s}\partial_{\xi_{ns}}\mathcal{W}_{n,\kappa_{s}}^{\left(\alpha_{s},1\right)}\left(\mu_{s},\xi_{ns}\right)\right]\\
\chi_{yz}^{\left(s\right)}= & \,i\frac{\omega_{ps}^{2}}{\omega^{2}}\frac{w_{\parallel s}}{w_{\perp s}}\frac{k_{\perp}w_{\perp s}}{2\Omega_{s}}\sum_{n\to-\infty}^{\infty}\left(\xi_{0s}-A_{s}\xi_{ns}\right)\nonumber \\
 & \times\partial_{\mu_{s},\xi_{ns}}^{2}\mathcal{Z}_{n,\kappa_{s}}^{\left(\alpha_{s},1\right)}\left(\mu_{s},\xi_{ns}\right)\\
\chi_{zz}^{\left(s\right)}= & -\frac{\omega_{ps}^{2}}{\omega^{2}}\frac{w_{\parallel s}^{2}}{w_{\perp s}^{2}}\sum_{n\to-\infty}^{\infty}\left(\xi_{0s}-A_{s}\xi_{ns}\right)\nonumber \\
 & \times\xi_{ns}\partial_{\xi_{ns}}\mathcal{Z}_{n,\kappa_{s}}^{\left(\alpha_{s},1\right)}\left(\mu_{s},\xi_{ns}\right),\label{eq:ST-BK-6}
\end{align}
where\end{subequations}
\[
\mathcal{W}_{n,\kappa}^{\left(\alpha,\beta\right)}\left(\mu,\xi\right)=\frac{n^{2}}{\mu}\mathcal{Z}_{n,\kappa}^{\left(\alpha,\beta\right)}\left(\mu,\xi\right)-2\mu\mathcal{Y}_{n,\kappa}^{\left(\alpha,\beta\right)}\left(\mu,\xi\right).
\]
Notice that the off-diagonal components of $\chi_{ij}$ obey the symmetry
relations $\chi_{xy}=-\chi_{yx}$, $\chi_{xz}=\chi_{zx}$, and $\chi_{yz}=-\chi_{yz}$.

In (\ref{eq:SuscTensor_Bi-Kappa}a-f) we have defined the parameters
$\mu_{s}=k_{\perp}^{2}\rho_{s}^{2}$, $\rho_{s}^{2}=w_{\perp s}^{2}/2\Omega_{s}^{2}$,
and $\xi_{ns}=\left(\omega-n\Omega_{s}\right)/k_{\parallel}w_{\parallel s}$\@.
The parameter $\rho_{s}$ is the (kappa modified) gyroradius (or Larmor
radius) of particle $s$\@. Hence, $\mu_{s}$ is the normalized gyroradius,
proportional to the ratio between $\rho_{s}$ and $\lambda_{\perp}$,
the perpendicular projection of the wavelength. The magnitude of $\mu_{s}$
quantifies the finite Larmor radii effects on wave propagation. On
the other hand, the parameter $\xi_{ns}$ quantifies the linear wave-particle
interactions in a finite-temperature plasma. Also in (\ref{eq:SuscTensor_Bi-Kappa}a-f),
the quantity 
\[
A_{s}=1-\frac{w_{\perp s}}{w_{\parallel s}}
\]
is the anisotropy parameter, which quantifies the effects of the VDF's
departure from an isotropic distribution, due to the temperature anisotropy.
The symbol $\partial_{z_{1},\dots,z_{n}}^{n}=\partial^{n}/\left(\partial z_{1}\cdots\partial z_{n}\right)$
is the $n$-th order partial derivative, relative to $z_{1},\dots,z_{n}$\@. 

Finally, the functions $\mathcal{Z}_{n,\kappa}^{\left(\alpha,\beta\right)}\left(\mu,\xi\right)$
and $\mathcal{Y}_{n,\kappa}^{\left(\alpha,\beta\right)}\left(\mu,\xi\right)$
are the so-called \emph{two-variables kappa plasma functions}. Their
definitions were first given in Ref. \onlinecite{GaelzerZiebell16/02}
(hereafter called Paper I) and are repeated in Eqs. (\ref{eq:GZM16:Z_cal-1_int-Z})-(\ref{eq:GZM16:Y_cal-1_int-H})\@.
Some properties and representations of $\mathcal{Z}$ and $\mathcal{Y}$
were also obtained in Paper I and several new properties and representations
will be derived in Sec. \ref{sec:GZM16:Kappa_plasma_functions}\@.
The evaluation of the functions $\mathcal{Z}$ and $\mathcal{Y}$
is determined not only by their arguments $\mu$ (normalized gyroradius)
and $\xi$ (wave-particle resonance), but also by a set of parameters:
$n$ (harmonic number), $\kappa$ (kappa index), and the pair $\left(\alpha,\beta\right)$\@.
Parameter $\alpha$ is the same real number adopted for the $\kappa$VDF
(\ref{eq:KAP7:f_s})\@. This parameter can be ignored and removed
from the equations if the distribution model is fixed. On the other
hand, the real parameter $\beta$ is crucial for the formalism. The
value of $\beta$ is related to the specific dielectric tensor component,
wave polarization and mathematical properties of the kappa plasma
functions.

The isotropic limit of $\chi_{ij}^{(s)}$ is obtained from (\ref{eq:SuscTensor_Bi-Kappa}a-f)
by setting $A_{s}=0$ (and $w_{\perp s}=w_{\parallel s}=w_{s}$)\@.
In this case the susceptibility tensor for each particle species reduces
to the form that can be easily gleaned from the Cartesian components
of $\varepsilon_{ij}$ presented in Appendix C of Paper I\@. On the
other hand, the susceptibility tensor of a bi-Maxwellian plasma is
also obtained from (\ref{eq:SuscTensor_Bi-Kappa}a-f) by the process
called the \emph{Maxwellian limit}, i.e., the result of taking the
limit $\kappa_{s}\to\infty$, for any species $s$\@. The Maxwellian
limit of $\chi_{ij}^{(s)}$ is given by Eqs. (\ref{eq:SuscTensor_Bi-Maxwellian}a-f).

Equations (\ref{eq:DielTensor-general}-\ref{eq:SuscTensor_Bi-Kappa})
show the general form for the dielectric tensor of a bi-Kappa plasma.
These expressions, along with the representations of the kappa plasma
functions derived in Paper I and in Sec. \ref{sec:GZM16:Kappa_plasma_functions},
contain sufficient information for a methodical study of the properties
of wave propagation and emission/absorption in an anisotropic, superthermal
plasma. Future works will implement an analysis of temperature-anisotropy-driven
instabilities excited in low-frequency parallel- and oblique-propagating
electromagnetic eigenmodes.

\section{New expressions for the kappa plasma special functions\label{sec:GZM16:Kappa_plasma_functions}}

\subsection{The superthermal plasma gyroradius function}

The function $\mathcal{H}_{n,\kappa}^{\left(\alpha,\beta\right)}\left(z\right)$
quantifies the physical effects on wave propagation due to the particles's
finite gyroradii when their probability distribution function is described
by a kappa VDF\@. For this reason, it was named by Paper I as the
(kappa) \emph{plasma gyroradius function }($\kappa$PGF)\@. The basic
definition of this function was given in Eq. (I.20) (i.e., Eq. 20
of Paper I) and is repeated here,
\begin{align}
\mathcal{H}_{n,\kappa}^{\left(\alpha,\beta\right)}\left(z\right) & =2\int_{0}^{\infty}dx\,\frac{xJ_{n}^{2}\left(yx\right)}{\left(1+x^{2}/\kappa\right)^{\lambda-1}}, & \left(y^{2}=2z\right),\label{eq:GZM16:Kappa-PGF}
\end{align}
where $\lambda=\kappa+\alpha+\beta$.

The Maxwellian limit of this function is the well-know representation
in terms of the modified Bessel function,\cite{OlverMaximon-NIST10}
\begin{equation}
\lim_{\kappa\to\infty}\mathcal{H}_{n,\kappa}^{\left(\alpha,\beta\right)}\left(z\right)=\mathscr{H}_{n}\left(z\right)=e^{-z}I_{n}\left(z\right).\label{eq:GZM16:Maxwellian-PGF}
\end{equation}

Sections III.B and A.2 of Paper I contain several mathematical properties
of $\mathcal{H}_{n,\kappa}^{\left(\alpha,\beta\right)}\left(z\right)$
and most of them will not be shown here, with a few important exceptions.
One of the exceptions is its general, closed-form representation in
terms of the Meijer $G$-function, as shown in Eq. (I.22)\@. Namely,\begin{subequations}\label{eq:GZM16:kPGF-G}
\begin{gather}
\mathcal{H}_{n,\kappa}^{\left(\alpha,\beta\right)}\left(z\right)=\frac{\pi^{-1/2}\kappa}{\Gamma\left(\lambda-1\right)}G_{1,3}^{2,1}\left[2\kappa z\left|{\nicefrac{1}{2}\atop \lambda-2,n,-n}\right.\right]\label{eq:GZM16:kPGF-G1}\\
=\frac{\pi^{-1/2}\kappa}{\Gamma\left(\lambda-1\right)}G_{3,1}^{1,2}\left[\frac{1}{2\kappa z}\left|{3-\lambda,1-n,1+n\atop \nicefrac{1}{2}}\right.\right].\label{eq:GZM16:kPGF-G2}
\end{gather}
\end{subequations}Representation (\ref{eq:GZM16:kPGF-G2}) was obtained
employing the symmetry property of the $G$-function given by Eq.
(I.11a).

The definition and some properties of the $G$-function can be found
in Sec. B.2 of Paper I and in the cited literature. Some additional
properties, employed in the present paper, are given in Appendix \ref{sec:GZM16:G-functions}.

Additional mathematical properties of the $\mathcal{H}$-function,
that were not included in Paper I, will be presented here.

\subsubsection{Derivatives}

Equations (I.25a)-(I.25d) show recurrence relations for the $\mathcal{H}$-function
that involve its first derivative and that in the Maxwellian limit
reduce to the respective relations for $\mathscr{H}_{n}\left(z\right)$,
easily obtained from the properties of the modified Bessel function.

It is also possible to obtain closed-form representations for the
derivatives of $\mathcal{H}_{n,\kappa}^{\left(\alpha,\beta\right)}\left(z\right)$
in any order. Applying the operator $D^{k}\equiv d^{k}/dz^{k}$ $\left(k=0,1,2,\dots\right)$
on (\ref{eq:GZM16:kPGF-G1}) and employing identity (\ref{eq:der1}),
we obtain\begin{subequations}\label{eq:GZM16:H-d(k)}
\begin{equation}
\frac{\mathcal{H}_{n,\kappa}^{\left(\alpha,\beta\right)\left(k\right)}\left(z\right)}{\left(-z\right)^{-k}}=\frac{\pi^{-1/2}\kappa}{\Gamma\left(\lambda-1\right)}G_{2,4}^{3,1}\left[2\kappa z\left|{\nicefrac{1}{2},0\atop k,\lambda-2,n,-n}\right.\right],\label{eq:GZM16:H-d(k)-G}
\end{equation}
where $\mathcal{H}^{\left(k\right)}=d^{k}\mathcal{H}/dz^{k}$\@. 

Formula (\ref{eq:GZM16:H-d(k)-G}) is valid for any $z$ and $k$,
but the value of $\mathcal{H}$ at the origin must be treated separately.
Applying the operator $D^{k}$ on the definition (\ref{eq:GZM16:Kappa-PGF}),
we can employ the power series expansion of $J_{n}^{2}\left(yx\right)$
given by Eq. (10.8.3) of Ref. \onlinecite{OlverMaximon-NIST10} in
order to evaluate the integral in the limit $y\to0$, thereby obtaining
\[
\frac{\mathcal{H}_{n,\kappa}^{\left(\alpha,\beta\right)\left(k\right)}\left(0\right)}{\left(2k\right)!\kappa}=\left(\frac{-\kappa}{2}\right)^{k}\frac{\left(\lambda-2\right)_{-k}}{\lambda-2}\sum_{\ell=0}^{k}\frac{\left(-\right)^{\ell}\delta_{\left|n\right|,\ell}}{\left(k+\ell\right)!\left(k-\ell\right)!},
\]
which is valid for $\lambda>2+k$\@. Here, $\delta_{n,m}$ is the
Kronecker delta and $\left(a\right)_{\ell}=\Gamma\left(a+\ell\right)/\Gamma\left(a\right)$
is the Pochhammer symbol. One can easily verify that the case $k=0$
reduces to Eq. (I.21).

As it happens with $\mathcal{H}_{n,\kappa}^{\left(\alpha,\beta\right)}\left(z\right)$,
its derivative in any order has two different representations in terms
of more usual functions, depending on whether $\lambda$ is integer
or not. These cases will now be addressed.

\paragraph{Case $\lambda$ noninteger.}
\begin{widetext}
If $\lambda\neq2,3,\dots$, then we can employ the representation
of the $G$-function in terms of generalized hypergeometric functions,
given by Eq. (I.B14)\@. Hence, we have
\begin{multline}
\frac{\mathcal{H}_{n,\kappa}^{\left(\alpha,\beta\right)\left(k\right)}\left(z\right)}{\left(-z\right)^{-k}}=\frac{\pi^{-1/2}\kappa}{\Gamma\left(\lambda-1\right)}\left[\frac{\Gamma\left(n+2-\lambda\right)\Gamma\left(\lambda-\nicefrac{3}{2}\right)}{\Gamma\left(\lambda-1+n\right)}\left(2-\lambda\right)_{k}\left(2\kappa z\right)^{\lambda-2}\pFq 23\left({\lambda-\nicefrac{3}{2},\lambda-1\atop \lambda-1-n,\lambda-1+n,\lambda-1-k};2\kappa z\right)\right.\\
\left.+\frac{\Gamma\left(\lambda-2-n\right)\Gamma\left(n+\nicefrac{1}{2}\right)}{\Gamma\left(2n+1\right)}\left(-n\right)_{k}\left(2\kappa z\right)^{n}\pFq 23\left({n+\nicefrac{1}{2},n+1\atop n+3-\lambda,2n+1,n+1-k};2\kappa z\right)\right],\label{eq:GZM16:H-d(k)-2F3}
\end{multline}
where $\pFq 23\bigl(\cdots;z\bigr)$ is another hypergeometric series
of class 1, discussed in Sec. B.1 of Paper I\@. The case $k=0$ reduces
to Eq. (I.23).
\end{widetext}

\paragraph{Case $\lambda$ integer.}

Now, writing $\lambda=m+2$ $\left(m=0,1,2,\dots\right)$ in (\ref{eq:GZM16:H-d(k)-G})
and looking at the representation (\ref{eq:GZM16:I_muK_nu-G}), we
notice that if we choose $\mu=n-k$ and $\nu=n+k$ and employ the
differentiation formula (I.B13a), we can write
\begin{multline*}
\mathcal{H}_{n,\kappa}^{\left(\alpha,\beta\right)\left(k\right)}\left(z\right)=\frac{2\kappa\left(-z\right)^{-k}\left(-2\kappa z\right)^{m}}{\Gamma\left(m+1\right)}\\
\times\left.\frac{d^{m+k}}{dy^{m+k}}\left[y^{k}I_{n-k}\left(\sqrt{y}\right)K_{n+k}\left(\sqrt{y}\right)\right]\right|_{y=2\kappa z},
\end{multline*}
where $K_{m}\left(z\right)$ is the second modified Bessel function\@.\cite{OlverMaximon-NIST10}
Finally, employing Leibniz formula for the derivative\cite{Roy+-NIST10}
and the identities written just before Eq. (I.24), we obtain
\begin{multline}
\mathcal{H}_{n,\kappa}^{\left(\alpha,\beta\right)\left(k\right)}\left(z\right)=\frac{2\kappa z^{k}}{\Gamma\left(m+1\right)}\left(\frac{\kappa z}{2}\right)^{\left(m+k\right)/2}\sum_{s=0}^{m+k}\left(-\right)^{s}\\
\times\binom{m+k}{s}K_{n-m+s}\left(\sqrt{2\kappa z}\right)I_{n-k+s}\left(\sqrt{2\kappa z}\right).\label{eq:GZM16:H-d(k)-IK}
\end{multline}
As expected, for $k=0$ this result reduces to (I.24).

\end{subequations}

\subsubsection{Asymptotic expansion}

The representation of the $\mathcal{H}$-function given by (\ref{eq:GZM16:kPGF-G2})
is formally exact for any $z$ and the function could be formally
expressed in terms of the $\pFq 30\left(\cdots;z\right)$ hypergeometric
series, after using Eq. (I.B14)\@. However, as explained in Sec.
B.1 of Paper I, the $\pFq 30$ belongs to class 3, whose series are
everywhere divergent, except at $z=0$\@. Hence, the representation
of $\mathcal{H}_{n,\kappa}^{\left(\alpha,\beta\right)}\left(z\right)$
in terms of $\pFq 30$ only makes sense when one is looking for an
asymptotic expansion of $\mathcal{H}$, which provides a finite number
of correct digits when $z\gg1$ if only a finite numbers of terms
in the series expansion is kept. 

With this caveat in mind, using Eq. (I.B14) in (\ref{eq:GZM16:kPGF-G2}),
we obtain
\begin{multline*}
\mathcal{H}_{n,\kappa}^{\left(\alpha,\beta\right)}\left(z\right)=\frac{1}{\sqrt{\pi}}\frac{\kappa\Gamma\left(\lambda-\nicefrac{3}{2}\right)}{\Gamma\left(\lambda-1\right)\sqrt{2\kappa z}}\\
\times\pFq 30\left({\lambda-\nicefrac{3}{2},\nicefrac{1}{2}+n,\nicefrac{1}{2}-n\atop -};\frac{1}{2\kappa z}\right),
\end{multline*}
which, as explained, is only valid on the limit $z\to\infty$\@.
Inserting the series (I.B1), we obtain the asymptotic expansion
\begin{multline}
\mathcal{H}_{n,\kappa}^{\left(\alpha,\beta\right)}\left(z\right)\simeq\frac{1}{\sqrt{\pi}}\frac{\kappa\Gamma\left(\lambda-\nicefrac{3}{2}\right)}{\Gamma\left(\lambda-1\right)\sqrt{2\kappa z}}\\
\times\sum_{k=0}\frac{\left(\lambda-\nicefrac{3}{2}\right)_{k}\left(\nicefrac{1}{2}+n\right)_{k}\left(\nicefrac{1}{2}-n\right)_{k}}{k!\left(2\kappa z\right)^{k}}.\label{eq:GZM16:H-asymptotic}
\end{multline}
Notice that the upper limit of the sum is absent. This upper limit
must be computationally determined, taking into account the desired
number of correct digits in the evaluation of $\mathcal{H}_{n,\kappa}^{\left(\alpha,\beta\right)}\left(z\right)$.

The Maxwellian limit of (\ref{eq:GZM16:H-asymptotic}) renders
\[
\mathscr{H}_{n}\left(z\right)\simeq\frac{1}{\sqrt{2\pi z}}\sum_{k=0}\frac{\left(-\right)^{k}\Gamma\left(n+k+\nicefrac{1}{2}\right)}{\Gamma\left(n-k+\nicefrac{1}{2}\right)k!\left(2z\right)^{k}},
\]
which is exactly the asymptotic expansion of $\mathscr{H}_{n}\left(z\right)$
given by Eq. (8.451.5) of Ref. \onlinecite{GradshteynRyzhik07}.

\subsubsection{Sum rule}

Sum rules are useful for the numerical evaluation of special functions.
If we sum (\ref{eq:GZM16:Kappa-PGF}) over all harmonic numbers and
use the identity\cite{OlverMaximon-NIST10} 
\begin{equation}
\sum_{n\to-\infty}^{\infty}J_{n}^{2}\left(z\right)=1,\label{eq:GZM16:BesselJ-sum}
\end{equation}
the remaining integral can be evaluated by the definition of the Beta
function,\cite{AskeyRoy-NIST10} resulting
\begin{equation}
\sum_{n\to-\infty}^{\infty}\mathcal{H}_{n,\kappa}^{\left(\alpha,\beta\right)}\left(z\right)=\frac{\kappa}{\lambda-2}.\label{eq:GZM16:H-sum_rule}
\end{equation}
Several other sum rules for $\mathcal{H}$ can be found in the same
fashion.

\subsubsection{The associated gyroradius function}

Among the representations for the two-variable functions $\mathcal{Z}_{n,\kappa}^{\left(\alpha,\beta\right)}\left(\mu,\xi\right)$
and $\mathcal{Y}_{n,\kappa}^{\left(\alpha,\beta\right)}\left(\mu,\xi\right)$,
derived in the section \ref{sub:GZM16:ZY_cal-Series-representations},
the following function appears,
\begin{equation}
\widetilde{\mathcal{H}}_{n,k,\kappa}^{\left(\alpha,\beta\right)}\left(\mu\right)=\frac{\pi^{-1/2}\kappa}{\Gamma\left(\lambda-1\right)}G_{1,3}^{2,1}\left[2\kappa\mu\left|{\nicefrac{1}{2}-k\atop \lambda-2,n,-n}\right.\right],\label{eq:GZM16:H-associated}
\end{equation}
which is clearly related to $\mathcal{H}_{n,\kappa}^{\left(\alpha,\beta\right)}\left(z\right)$,
differing by the parameter $k$\@. For this reason, it is christened
here as the \emph{associated plasma gyroradius function.}

Some properties of the $\widetilde{\mathcal{H}}$-function are now
presented. A trivial property is $\widetilde{\mathcal{H}}_{n,0,\kappa}^{\left(\alpha,\beta\right)}\left(z\right)=\mathcal{H}_{n,\kappa}^{\left(\alpha,\beta\right)}\left(z\right)$.

\paragraph{Relation with $\mathcal{H}_{n,\kappa}^{\left(\alpha,\beta\right)}\left(z\right)$\@.}

The associated PGF is related to the $\mathcal{H}$-function and its
derivatives. First, due to the differentiation formula (\ref{eq:der2}),
it is clear that we can write
\[
\widetilde{\mathcal{H}}_{n,k,\kappa}^{\left(\alpha,\beta\right)}\left(\mu\right)=\mu^{1/2}\frac{d^{k}}{d\mu^{k}}\left[\mu^{k-1/2}\mathcal{H}_{n,\kappa}^{\left(\alpha,\beta\right)}\left(\mu\right)\right].
\]
Then, using Leibniz's formula and the formula for $D^{m}z^{\gamma}$
just above (\ref{eq:der2}), we obtain\begin{subequations}\label{eq:GZM16:H_k-dH-RecRel}
\begin{equation}
\widetilde{\mathcal{H}}_{n,k,\kappa}^{\left(\alpha,\beta\right)}\left(\mu\right)=\Gamma\left(k+\frac{1}{2}\right)\sum_{\ell=0}^{k}\binom{k}{\ell}\frac{\mu^{\ell}\mathcal{H}_{n,\kappa}^{\left(\alpha,\beta\right)\left(\ell\right)}\left(\mu\right)}{\Gamma\left(\ell+\nicefrac{1}{2}\right)}.\label{eq:GZM16:dH->H_k-RR}
\end{equation}

The reciprocate relation is obtained starting from (\ref{eq:GZM16:H-d(k)-G}),
which is written in terms of the Mellin-Barnes integral with the help
of (I.B10)\@. Then, we have
\begin{multline*}
\mathcal{H}_{n,\kappa}^{\left(\alpha,\beta\right)\left(k\right)}\left(\mu\right)=\frac{\kappa\left(-\mu\right)^{-k}}{2\pi^{3/2}i\Gamma\left(\lambda-1\right)}\\
\times\int_{L}\frac{\Gamma\left(\lambda-2-s\right)\Gamma\left(n-s\right)\Gamma\left(\nicefrac{1}{2}+s\right)}{\Gamma\left(n+1+s\right)\left(2\kappa\mu\right)^{-s}}\left(-s\right)_{k}ds.
\end{multline*}
On the other hand, from (\ref{eq:GZM16:H-associated}) and (I.B10)
again, we have
\begin{multline*}
\widetilde{\mathcal{H}}_{n,k,\kappa}^{\left(\alpha,\beta\right)}\left(\mu\right)=\frac{\left(2\pi^{3/2}i\right)^{-1}\kappa}{\Gamma\left(\lambda-1\right)}\\
\times\int_{L}\frac{\Gamma\left(\lambda-2-s\right)\Gamma\left(n-s\right)\Gamma\left(\nicefrac{1}{2}+s\right)}{\Gamma\left(1+n+s\right)\left(2\kappa\mu\right)^{-s}}\left(\frac{1}{2}+s\right)_{k}ds.
\end{multline*}
Then, if we employ the identity 
\[
\left(a+b\right)_{n}=\sum_{\ell=0}^{n}\left(-\right)^{\ell}\binom{n}{\ell}\left(a+\ell\right)_{n-\ell}\left(-b\right)_{\ell},
\]
we can finally write the reciprocate relation 
\begin{equation}
\frac{\mathcal{H}_{n,\kappa}^{\left(\alpha,\beta\right)\left(k\right)}\left(\mu\right)}{\mu^{-k}}=\sum_{\ell=0}^{k}\binom{k}{\ell}\left(\frac{1}{2}-k\right)_{k-\ell}\widetilde{\mathcal{H}}_{n,\ell,\kappa}^{\left(\alpha,\beta\right)}\left(\mu\right).\label{eq:GZM16:H_k->dH-RR}
\end{equation}
\end{subequations}

\paragraph{Representations.}

The computation of $\widetilde{\mathcal{H}}_{n,k,\kappa}^{\left(\alpha,\beta\right)}\left(\mu\right)$
can be carried out as follows. For noninteger $\lambda$, it is more
efficient to employ identity (I.B14) and evaluate
\begin{multline}
\widetilde{\mathcal{H}}_{n,k,\kappa}^{\left(\alpha,\beta\right)}\left(z\right)=\frac{\pi^{-1/2}\kappa}{\Gamma\left(\lambda-1\right)}\left[\frac{\Gamma\left(n+2-\lambda\right)}{\Gamma\left(\lambda-1+n\right)}\left(2\kappa z\right)^{\lambda-2}h_{k}\left(z\right)\right.\\
\left.+\frac{\Gamma\left(\lambda-2-n\right)}{\Gamma\left(2n+1\right)}\left(2\kappa z\right)^{n}g_{k}\left(z\right)\right],\label{eq:GZM16:H_k-1F2}
\end{multline}
where
\begin{align*}
\frac{h_{k}\left(z\right)}{\Gamma\left(\lambda-\nicefrac{3}{2}+k\right)}= & \pFq 12\left({\lambda-\nicefrac{3}{2}+k\atop \lambda-1-n,\lambda-1+n};2\kappa z\right)\\
\frac{g_{k}\left(z\right)}{\Gamma\left(n+\nicefrac{1}{2}+k\right)}= & \pFq 12\left({n+\nicefrac{1}{2}+k\atop n+3-\lambda,2n+1};2\kappa z\right).
\end{align*}

On the other hand, for integer $\lambda$ the only representation
found for $\widetilde{\mathcal{H}}$  similar to (\ref{eq:GZM16:H-d(k)-IK})
contains a double sum. Consequently, it is equivalent to simply employ
Eqs. (\ref{eq:GZM16:dH->H_k-RR}) and (\ref{eq:GZM16:H-d(k)-IK}).

\paragraph{Recurrence relation.}

The numerical computation of $\widetilde{\mathcal{H}}_{n,k,\kappa}^{\left(\alpha,\beta\right)}\left(z\right)$
can be carried out using either Eq. (\ref{eq:GZM16:H_k-1F2}) or Eqs.
(\ref{eq:GZM16:dH->H_k-RR}) and (\ref{eq:GZM16:H-d(k)-IK}) combined.
However, since the associate function appears in series involving
the parameter $k$, if a recurrence relation for $\widetilde{\mathcal{H}}_{n,k,\kappa}^{\left(\alpha,\beta\right)}\left(z\right)\equiv\widetilde{\mathcal{H}}\left[k\right]$
on this parameter could be found, it could substantially reduce the
computational time required for the evaluation of the series.

\begin{widetext}

Such recurrence relation can be found by first considering the particular
case of noninteger $\lambda$, given by Eq. (\ref{eq:GZM16:H_k-1F2})\@.
We observe that the auxiliary functions $h_{k}\left(z\right)$ and
$g_{k}\left(z\right)$ in (\ref{eq:GZM16:H_k-1F2}) and, consequently,
the function $\widetilde{\mathcal{H}}\left[k\right]$ itself, all
obey the same four-term recurrence relation, which can be derived
from the corresponding relation for the function $\pFq 12\left(\cdots;z\right)$
in the upper parameter, given by Ref. \onlinecite{1F2-RR-1}\@. Namely,
\begin{multline}
\widetilde{\mathcal{H}}\left[k+3\right]-\left(\lambda+\frac{5}{2}+3k\right)\widetilde{\mathcal{H}}\left[k+2\right]+\left[2\lambda-n^{2}-\frac{3}{4}+2\left(\lambda+1+\frac{3}{2}k\right)k-2\kappa z\right]\widetilde{\mathcal{H}}\left[k+1\right]\\
+\left(\lambda-\frac{3}{2}+k\right)\left(n+\frac{1}{2}+k\right)\left(n-\frac{1}{2}-k\right)\widetilde{\mathcal{H}}\left[k\right]=0.\label{eq:GZM16:H_k-RR}
\end{multline}

Although the relation (\ref{eq:GZM16:H_k-RR}) was derived for noninteger
$\lambda$, it can be easily shown that it is indeed valid for any
$\lambda$\@. Substituting into the functions $\widetilde{\mathcal{H}}\left[k\right]$
in (\ref{eq:GZM16:H_k-RR}) the definition (\ref{eq:GZM16:H-associated})
and then the corresponding representations in terms of Mellin-Barnes
integrals (Eq. I.B10), one can show, by using known properties of
the gamma function,\cite{AskeyRoy-NIST10} that the identity (\ref{eq:GZM16:H_k-RR})
is indeed valid for any real $\lambda$.

\end{widetext}

\subsection{The superthermal plasma dispersion function}

The superthermal (or kappa) \emph{plasma dispersion function} ($\kappa$PDF)
was defined by Eq. (I.11), and several of its properties were discussed
in sections III.A and A.1 of Paper I\@. Here, we will merely present
a few additional properties, which were not included in Paper I and
are important for the work at hand.

\subsubsection{Representations in terms of the $\boldsymbol{G}$-function.}

Taking the representations (I.15) for $Z_{\kappa}^{\left(\alpha,\beta\right)}\left(\xi\right)$
and (I.B15a) for the Gauss function, we have\begin{subequations}\label{eq:GZM16:kPDF-G}
\begin{multline}
Z_{\kappa}^{\left(\alpha,\beta\right)}\left(\xi\right)=-\frac{\pi^{1/2}\kappa^{-\beta-1}\xi}{\Gamma\left(\sigma-\nicefrac{3}{2}\right)}G_{2,2}^{1,2}\left[\frac{\xi^{2}}{\kappa}\left|{0,\nicefrac{3}{2}-\lambda\atop 0,-\nicefrac{1}{2}}\right.\right]\\
+\frac{i\pi^{1/2}\Gamma\left(\lambda-1\right)}{\kappa^{\beta+1/2}\Gamma\left(\sigma-\nicefrac{3}{2}\right)}\left(1+\frac{\xi^{2}}{\kappa}\right)^{-\left(\lambda-1\right)}.\label{eq:GZM16:kPDF-G-1}
\end{multline}
As explained in Paper I, the Maxwellian limit of this representation
reduces to the known expression of the Fried \& Conte function in
terms of the Kummer confluent hypergeometric series.

Another, more compact, representation is obtained if we first modify
the limits of the integral in (I.11) to the interval $0\leqslant s<\infty$,
define the new integration variable $s=\sqrt{u}$ and identify the
resulting integration with the identity (I.B12)\@. Proceeding in
this way, we obtain the equivalent representation
\begin{equation}
Z_{\kappa}^{\left(\alpha,\beta\right)}\left(\xi\right)=\frac{\pi^{-1/2}\kappa^{-\beta-1}\xi}{\Gamma\left(\sigma-\nicefrac{3}{2}\right)}G_{2,2}^{2,2}\left[-\frac{\xi^{2}}{\kappa}\left|{0,\nicefrac{3}{2}-\lambda\atop 0,-\nicefrac{1}{2}}\right.\right].\label{eq:GZM16:kPDF-G-2}
\end{equation}

Taking the limit $\kappa\to\infty$ of (\ref{eq:GZM16:kPDF-G-2}),
and identifying the result with (\ref{eq:GZM16:U-G}), we obtain
\[
\lim_{\kappa\to\infty}Z_{\kappa}^{\left(\alpha,\beta\right)}\left(\xi\right)=\xi U\left({1\atop \nicefrac{3}{2}};-\xi^{2}\right),
\]
where $U\left(\cdots;z\right)$ is the Tricomi confluent hypergeometric
function\@.\cite{Daalhuis-NIST10a} This is another known representation
of the Fried \& Conte function.\cite{Peratt84/03}

\end{subequations}

\subsubsection{The associated plasma dispersion function}

The \emph{associated plasma dispersion function}, defined by
\begin{multline}
\widetilde{Z}_{k,\kappa}^{\left(\alpha,\beta\right)}\left(\xi\right)\doteq\frac{\kappa^{-\left(k+\beta+1/2\right)}\Gamma\left(\lambda-1\right)}{\sqrt{\pi}\Gamma\left(\sigma-\nicefrac{3}{2}\right)}\\
\times\int_{-\infty}^{\infty}ds\,\frac{s^{2k}\left(1+s^{2}/\kappa\right)^{-\left(\lambda-\nicefrac{3}{2}+k\right)}}{s-\xi},\label{eq:GZM16:Z-associated}
\end{multline}
is another new special function that appears in the series expansions
derived in section \ref{sub:GZM16:ZY_cal-Series-representations}
for the two-variables special functions $\mathcal{Z}_{n,\kappa}^{\left(\alpha,\beta\right)}\left(\mu,\xi\right)$
and $\mathcal{Y}_{n,\kappa}^{\left(\alpha,\beta\right)}\left(\mu,\xi\right)$\@.
It has the trivial property 
\begin{equation}
\widetilde{Z}_{0,\kappa}^{\left(\alpha,\beta\right)}\left(\xi\right)=\frac{\Gamma\left(\lambda-1\right)}{\sqrt{\kappa}\Gamma\left(\lambda-\nicefrac{3}{2}\right)}Z_{\kappa}^{\left(\alpha,\beta-\nicefrac{1}{2}\right)}\left(\xi\right),\label{eq:GZM16:Z_tilde-k0}
\end{equation}
and, in the following, alternative representations for the $\widetilde{Z}$-function
will be derived.

\paragraph{Representations.}

The first expression is valid when $\lambda$ is half-integer $\left(\lambda=\nicefrac{5}{2},\nicefrac{7}{2},\dots\right)$\@.
In this case, writing $m=\lambda-\nicefrac{3}{2}+k$ $\left(m=1,2,\dots\right)$,
all singular points in (\ref{eq:GZM16:Z-associated}), at $s=\xi$
and $s=\pm i\sqrt{\kappa}$, are poles and thus we are permitted to
evaluate $\widetilde{Z}_{k,\kappa}^{\left(\alpha,\beta\right)}\left(\xi\right)$
using the residue theorem, exactly as was done by Summers and Thorne\@.\cite{SummersThorne91/08}

Let us consider the contour integral
\[
I_{B}=\int_{B}ds\,\frac{s^{2k}\left(1+s^{2}/\kappa\right)^{-\left(\lambda+k-\nicefrac{3}{2}\right)}}{s-\xi},
\]
where the contour $B$ is comprised by the semicircle in the lower-half
plane of complex $s$ (with radius $S\to\infty$), which is closed
by the integration along the real line of $s$, deformed according
to the Landau prescription (i.e., circulating around the pole at $s=\xi$
from below)\@. See, for instance, the contour in Fig. 2 of Ref. \onlinecite{SummersThorne91/08},
but with closing in the lower-half $s$-plane. Then, it is easy to
show that the contribution along the semicircle of radius $S$ vanishes
as $S\to\infty$ and $I_{B}$ is simply evaluated from the residue
at $s=-i\sqrt{\kappa}$ as $I_{B}=-2\pi i\mathrm{Res}\left(-i\sqrt{\kappa}\right)$,
since the pole at $s=\xi$ is always outside $B$\@. 

The residue is evaluated by the usual formula for a pole of order
$m$,\cite{AblowitzFokas03} leading to the representation\begin{subequations}\label{eq:GZM16:Z_tilde-representations}
\begin{multline}
\widetilde{Z}_{k,\kappa}^{\left(\alpha,\beta\right)}\left(\xi\right)=\frac{2\sqrt{\pi}i\left(-\right)^{k}\Gamma\left(\lambda-1\right)}{\kappa^{\beta+1/2}\Gamma\left(\sigma-\nicefrac{3}{2}\right)}\\
\times\sum_{\ell=0}^{M}\sum_{r=0}^{m-1-\ell}\frac{\left(-2k\right)_{\ell}\left(m\right)_{r}\left(1\right)_{m-1-\ell-r}}{2^{m+r}\Gamma\left(m-\ell-r\right)\ell!r!}\\
\times\left(1-\frac{i\xi}{\sqrt{\kappa}}\right)^{-\left(m-\ell-r\right)},\label{eq:GZM16:Z_tilde-lambda-half-integer}
\end{multline}
where $M=\min\left(m-1,2k\right)$\@. One can easily verify in (\ref{eq:GZM16:Z_tilde-lambda-half-integer})
that for integer $\kappa$, 
\[
\widetilde{Z}_{0,\kappa}^{\left(1,\nicefrac{3}{2}\right)}\left(\xi\right)=\frac{\Gamma\left(\kappa+\nicefrac{3}{2}\right)}{\kappa^{1/2}\kappa!}Z_{\kappa}^{*}\left(\xi\right),
\]
where $Z_{\kappa}^{*}\left(\xi\right)$ is given by Eq. (20) of Ref.
\onlinecite{SummersThorne91/08}.

A different expression for the $\widetilde{Z}$-function will now
be obtained, which is valid for any real $\lambda$\@. We already
know that for $k=0$ the $\widetilde{Z}$-function is given in terms
of the $\kappa$PDF by (\ref{eq:GZM16:Z_tilde-k0})\@. Now, for $k\geqslant1$,
we modify the limits of the integral in (\ref{eq:GZM16:Z-associated})
to the interval $0\leqslant s<\infty$, define the new variable $s=\sqrt{u}$
and employ identity (I.B12) in order to obtain
\begin{multline}
\widetilde{Z}_{k,\kappa}^{\left(\alpha,\beta\right)}\left(\xi\right)=\frac{i\kappa^{-\left(\beta+1/2\right)}\Gamma\left(\lambda-1\right)}{\sqrt{\pi}\Gamma\left(\sigma-\nicefrac{3}{2}\right)\Gamma\left(\lambda+k-\nicefrac{3}{2}\right)}\\
\times G_{2,2}^{2,2}\left[-\frac{\xi^{2}}{\kappa}\left|{\nicefrac{1}{2},\nicefrac{5}{2}-\lambda\atop k,\nicefrac{1}{2}}\right.\right],\label{eq:GZM16:Z_tilde-G}
\end{multline}
which is a $G$-function representation of the associated PDF\@.
If we now employ formula (I.B13a), we can write
\begin{multline*}
\widetilde{Z}_{k,\kappa}^{\left(\alpha,\beta\right)}\left(\xi\right)=\frac{i\kappa^{-\left(k+\beta+1/2\right)}\Gamma\left(\lambda-1\right)}{\sqrt{\pi}\Gamma\left(\sigma-\nicefrac{3}{2}\right)\Gamma\left(\lambda+k-\nicefrac{3}{2}\right)}\\
\times\xi^{2k}\frac{d^{k}}{dz^{k}}G_{2,2}^{2,2}\left[z\left|{\nicefrac{1}{2},\nicefrac{5}{2}-\lambda\atop 0,\nicefrac{1}{2}}\right.\right],
\end{multline*}
where we have provisionally defined $z=-\xi^{2}/\kappa$\@. This
result will now be identified with the derivatives of $Z_{\kappa}^{\left(\alpha,\beta\right)}\left(\xi\right)$.

If we take representation (\ref{eq:GZM16:kPDF-G-2}) of the $\kappa$PDF,
evaluate the $k$-th derivative on $\xi$ and employ the differentiation
formula (Eq. 1.1.1.2 of Ref. \onlinecite{Brychkov08}), 
\begin{multline*}
\frac{d^{k}}{dz^{k}}\left[f\left(\sqrt{z}\right)\right]=\sum_{\ell=0}^{k-1}\left(-\right)^{\ell}\frac{\Gamma\left(k+\ell\right)}{\Gamma\left(k-\ell\right)\ell!}\\
\times\left(2\sqrt{z}\right)^{-k-\ell}f^{\left(k-\ell\right)}\left(\sqrt{z}\right),
\end{multline*}
which is valid for $k\geqslant1$, we finally obtain 
\begin{multline}
\widetilde{Z}_{k,\kappa}^{\left(\alpha,\beta\right)}\left(\xi\right)=\frac{\kappa^{-1/2}\Gamma\left(\lambda-1\right)}{2^{k}\Gamma\left(\lambda-\nicefrac{3}{2}+k\right)}\sum_{\ell=0}^{k-1}\frac{\Gamma\left(k+\ell\right)}{2^{\ell}\Gamma\left(k-\ell\right)\ell!}\\
\times\left(-\xi\right)^{k-\ell}Z_{\kappa}^{\left(\alpha,\beta-\nicefrac{1}{2}\right)\left(k-\ell\right)}\left(\xi\right).\label{eq:GZM16:Z_tilde-Z-k>=00003D1}
\end{multline}

A final representation for $\widetilde{Z}_{k,\kappa}^{\left(\alpha,\beta\right)}\left(\xi\right)$
will be derived by returning to (\ref{eq:GZM16:Z-associated}), changing
the integration variable to $t$, defined as $s^{2}=\kappa t^{-1}\left(1-t\right)$,
and comparing the resulting integral with the formula (I.B5)\@. In
this way, we obtain
\begin{multline}
\widetilde{Z}_{k,\kappa}^{\left(\alpha,\beta\right)}\left(\xi\right)=\frac{\Gamma\left(\lambda-1\right)B\left(\lambda-1,k+\nicefrac{1}{2}\right)}{\sqrt{\pi}\kappa^{\beta+1}\Gamma\left(\sigma-\nicefrac{3}{2}\right)}\\
\times\xi\pFq 21\left({1,\lambda-1\atop \lambda-\nicefrac{1}{2}+k};1+\frac{\xi^{2}}{\kappa}\right),\:\left(\Im\xi>0\right),\label{eq:eq:GZM16:Z_tilde-2F1}
\end{multline}
where $B\left(a,b\right)=\Gamma\left(a\right)\Gamma\left(b\right)/\Gamma\left(a+b\right)$
is the beta function\cite{AskeyRoy-NIST10} and $\pFq 21\left(\cdots;z\right)$
is the Gauss hypergeometric function\cite{Daalhuis-NIST10b} (see
also Sec. B.1 of Paper I)\@. It must be pointed out that the representation
(\ref{eq:eq:GZM16:Z_tilde-2F1}) is only valid for the upper-half
of the $\xi$-plane. In order to employ this expression when $\Im\xi\leqslant0$,
one must evaluate also its analytical continuation, employing the
same technique applied to Eq. (I.13)\@. The resulting expressions
for the functions $\mathcal{Z}$ and $\mathcal{Y}$ are shown in Eqs.
(\ref{eq:GZM16:Z_cal-HZ_tilde-2F1}) and (\ref{eq:GZM16:Y_cal-HZ_tilde-Gauss}).

\end{subequations}

\paragraph{Recurrence relation.}

The representation (\ref{eq:eq:GZM16:Z_tilde-2F1}) also allowed us
to obtain a recurrence relation for the associated PDF on the parameter
$k$\@. Employing the shorthand notation $\widetilde{Z}_{k,\kappa}^{\left(\alpha,\beta\right)}\left(\xi\right)\equiv\widetilde{Z}\left[k\right]$,
we can write
\begin{align*}
\widetilde{Z}\left[k\right] & =\frac{\left[\Gamma\left(\lambda-1\right)\right]^{2}}{\sqrt{\pi}\kappa^{\beta+1}\Gamma\left(\sigma-\nicefrac{3}{2}\right)}\xi z_{k}\left(\xi\right),\\
z_{k} & \doteq\frac{\Gamma\left(k+\nicefrac{1}{2}\right)}{\Gamma\left(\lambda-\nicefrac{1}{2}+k\right)}\pFq 21\left({1,\lambda-1\atop \lambda-\nicefrac{1}{2}+k};1+\frac{\xi^{2}}{\kappa}\right).
\end{align*}
Hence, if one finds the recurrence relation for the auxiliary function
$z_{k}\left(\xi\right)$, the same relation applies to $\widetilde{Z}\left[k\right]$\@.

Such a recurrence relation on the lower parameter of the Gauss function
is given by Ref. \onlinecite{2F1-RR-1}\@. Consequently, we obtain
\begin{multline}
\left(\lambda-\frac{1}{2}+k\right)\left(1+\frac{\xi^{2}}{\kappa}\right)\widetilde{Z}\left[k+2\right]\\
-\left[\left(\lambda-\frac{1}{2}+k\right)\frac{\xi^{2}}{\kappa}+\left(k+\frac{1}{2}\right)\left(1+\frac{\xi^{2}}{\kappa}\right)\right]\widetilde{Z}\left[k+1\right]\\
+\left(k+\frac{1}{2}\right)\frac{\xi^{2}}{\kappa}\widetilde{Z}\left[k\right]=0.\label{eq:GZM16:Z_tilde-RR}
\end{multline}
This result can be verified by inserting the definition (\ref{eq:GZM16:Z-associated})
in the place of $\widetilde{Z}\left[k\right]$ and then adequately
manipulating the integrand.

The three-term recurrence relation (\ref{eq:GZM16:Z_tilde-RR}) can
potentially reduce the computational time for the evaluation of the
functions $\mathcal{Z}_{n,\kappa}^{\left(\alpha,\beta\right)}\left(\mu,\xi\right)$
and $\mathcal{Y}_{n,\kappa}^{\left(\alpha,\beta\right)}\left(\mu,\xi\right)$,
discussed in the next section.

\subsection{The two-variables kappa plasma functions}

The dielectric tensor of a superthermal (kappa) plasma is written
in terms of the special functions $\mathcal{Z}_{n,\kappa}^{\left(\alpha,\beta\right)}\left(\mu,\xi\right)$
and $\mathcal{Y}_{n,\kappa}^{\left(\alpha,\beta\right)}\left(\mu,\xi\right)$,
collectively called the \emph{two-variables kappa plasma functions}
(2VKPs), as can be verified in Eqs. (I.6a-6d), for an isotropic $\kappa$VDF,
or in Eqs. (\ref{eq:ST-BK-1})-(\ref{eq:ST-BK-6}), for a bi-kappa
distribution.

The functions $\mathcal{Z}$ and $\mathcal{Y}$ were defined in Eqs.
(I.26a-26b) in terms of a single integral involving the superthermal
plasma dispersion function ($\kappa$PDF) $Z_{\kappa}^{\left(\alpha,\beta\right)}\left(\xi\right)$
(see Sec. III.A of Paper I)\@. These definitions will be repeated
below. We will include equivalent definitions in terms of double integrals,
which will also be used in this work.

\begin{widetext}

Hence, we define\begin{subequations}\label{eq:GZM16:ZY_cal}
\begin{align}
\mathcal{Z}_{n,\kappa}^{\left(\alpha,\beta\right)}\left(\mu,\xi\right) & =2\int_{0}^{\infty}dx\,\frac{xJ_{n}^{2}\left(\nu x\right)}{\left(1+x^{2}/\kappa\right)^{\lambda-1}}Z_{\kappa}^{\left(\alpha,\beta\right)}\left(\frac{\xi}{\sqrt{1+x^{2}/\kappa}}\right)\label{eq:GZM16:Z_cal-1_int-Z}\\
 & =\frac{2}{\pi^{1/2}\kappa^{1/2+\beta}}\frac{\Gamma\left(\lambda-1\right)}{\Gamma\left(\sigma-\nicefrac{3}{2}\right)}\int_{0}^{\infty}dx\,\int_{-\infty}^{\infty}ds\,\frac{xJ_{n}^{2}\left(\nu x\right)}{s-\xi}\left(1+\frac{x^{2}}{\kappa}+\frac{s^{2}}{\kappa}\right)^{-\left(\lambda-1\right)},\label{eq:GZM16:Z_cal-2_ints}\\
\mathcal{Y}_{n,\kappa}^{\left(\alpha,\beta\right)}\left(\mu,\xi\right) & =\frac{2}{\mu}\int_{0}^{\infty}dx\,\frac{x^{3}J_{n-1}\left(\nu x\right)J_{n+1}\left(\nu x\right)}{\left(1+x^{2}/\kappa\right)^{\lambda-1}}Z_{\kappa}^{\left(\alpha,\beta\right)}\left(\frac{\xi}{\sqrt{1+x^{2}/\kappa}}\right)\label{eq:GZM16:Y_cal-1_int-Z}\\
 & =\frac{2}{\pi^{1/2}\kappa^{1/2+\beta}\mu}\frac{\Gamma\left(\lambda-1\right)}{\Gamma\left(\sigma-\nicefrac{3}{2}\right)}\int_{0}^{\infty}dx\,\int_{-\infty}^{\infty}ds\,\frac{x^{3}J_{n-1}\left(\nu x\right)J_{n+1}\left(\nu x\right)}{s-\xi}\left(1+\frac{x^{2}}{\kappa}+\frac{s^{2}}{\kappa}\right)^{-\left(\lambda-1\right)},\label{eq:GZM16:Y_cal-2_ints}
\end{align}
where $\nu^{2}=2\mu$ and, as usual, $\sigma=\kappa+\alpha$ and $\lambda=\sigma+\beta$.

Other definitions in terms of a single integral can be obtained, which
are the counterparts of Eqs. (\ref{eq:GZM16:Z_cal-1_int-Z}) and (\ref{eq:GZM16:Y_cal-1_int-Z})\@.
If we change the order of the integrations in (\ref{eq:GZM16:Z_cal-2_ints})
and (\ref{eq:GZM16:Y_cal-2_ints}) and define a new integration variable
by $x=\sqrt{\chi}t$, where $\chi=1+s^{2}/\kappa$, the integral in
$t$ can be identified with (\ref{eq:GZM16:Kappa-PGF}) and we can
write
\begin{align}
\mathcal{Z}_{n,\kappa}^{\left(\alpha,\beta\right)}\left(\mu,\xi\right) & =\frac{\pi^{-1/2}}{\kappa^{\beta+1/2}}\frac{\Gamma\left(\lambda-1\right)}{\Gamma\left(\sigma-\nicefrac{3}{2}\right)}\int_{-\infty}^{\infty}ds\,\frac{\left(1+s^{2}/\kappa\right)^{-\left(\lambda-2\right)}}{s-\xi}\mathcal{H}_{n,\kappa}^{\left(\alpha,\beta\right)}\left[\mu\left(1+\frac{s^{2}}{\kappa}\right)\right]\label{eq:GZM16:Z_cal-1_int-H}\\
\mathcal{Y}_{n,\kappa}^{\left(\alpha,\beta\right)}\left(\mu,\xi\right) & =\frac{\pi^{-1/2}}{\kappa^{\beta-1/2}}\frac{\Gamma\left(\lambda-2\right)}{\Gamma\left(\sigma-\nicefrac{3}{2}\right)}\int_{-\infty}^{\infty}ds\,\frac{\left(1+s^{2}/\kappa\right)^{-\left(\lambda-4\right)}}{s-\xi}\mathcal{H}_{n,\kappa}^{\left(\alpha,\beta-1\right)\prime}\left[\mu\left(1+\frac{s^{2}}{\kappa}\right)\right].\label{eq:GZM16:Y_cal-1_int-H}
\end{align}
\end{subequations}

\end{widetext}

The Maxwellian limits of the 2VKPs was already obtained in Eq. (I.7)
and are
\begin{equation}
\begin{aligned}\lim_{\kappa\to\infty}\mathcal{Z}_{n,\kappa}^{\left(\alpha,\beta\right)}\left(\mu,\xi\right) & =\mathscr{H}_{n}\left(\mu\right)Z\left(\xi\right)\\
\lim_{\kappa\to\infty}\mathcal{Y}_{n,\kappa}^{\left(\alpha,\beta\right)}\left(\mu,\xi\right) & =\mathscr{H}_{n}'(\mu)Z\left(\xi\right),
\end{aligned}
\label{eq:GZM16:ZY_cal-Maxwellian}
\end{equation}
where the function $\mathscr{H}_{n}\left(\mu\right)$ is given by
(\ref{eq:GZM16:Maxwellian-PGF}) and $Z\left(\xi\right)$ is the usual
Fried \& Conte function, given, for instance, by Eq. (I.10).

Some new properties and representations of the functions $\mathcal{Z}$
and $\mathcal{Y}$ that were not included in Paper I will now be discussed.

\subsubsection{Derivatives of $\boldsymbol{\mathcal{Z}_{n,\kappa}^{\left(\alpha,\beta\right)}\left(\mu,\xi\right)}$}

As can be seen in Eqs. (\ref{eq:ST-BK-1})-(\ref{eq:ST-BK-6}), almost
all tensor components are given in terms of partial derivatives of
the function $\mathcal{Z}_{n,\kappa}^{\left(\alpha,\beta\right)}\left(\mu,\xi\right)$\@.
These derivatives can be easily computed from the direct function,
if one uses relations derived from the definitions (\ref{eq:GZM16:ZY_cal}).\begin{subequations}\label{eq:GZM16:Z_cal-ders}

We need the partial derivatives $\partial_{\xi}\mathcal{Z}$, $\partial_{\mu}\mathcal{Z}$,
and the mixed derivative $\partial_{\xi,\mu}^{2}\mathcal{Z}$\@.
Applying $\partial_{\xi}$ on (\ref{eq:GZM16:Z_cal-1_int-Z}) and
using Eq. (I.18a), we can identify with (\ref{eq:GZM16:Kappa-PGF})
and (\ref{eq:GZM16:Z_cal-1_int-Z}) and write
\begin{multline}
\partial_{\xi}\mathcal{Z}_{n,\kappa}^{\left(\alpha,\beta\right)}\left(\mu,\xi\right)=-2\left[\frac{\Gamma\left(\lambda-\nicefrac{1}{2}\right)}{\kappa^{\beta+1}\Gamma\left(\sigma-\nicefrac{3}{2}\right)}\mathcal{H}_{n,\kappa}^{\left(\alpha,\beta+\nicefrac{1}{2}\right)}\left(\mu\right)\right.\\
\left.\vphantom{\frac{\Gamma\left(\lambda-\nicefrac{1}{2}\right)}{\kappa^{\beta+1}\Gamma\left(\sigma-\nicefrac{3}{2}\right)}}+\xi\mathcal{Z}_{n,\kappa}^{\left(\alpha,\beta+1\right)}\left(\mu,\xi\right)\right].\label{eq:GZM16:dZ_cal-dxi}
\end{multline}

Now, applying $\partial_{\mu}$ on (\ref{eq:GZM16:Z_cal-2_ints})
and integrating by parts the $x$-integral, the resulting expression
can be manipulated in order to provide the relation between the derivatives
\begin{multline*}
\mu\partial_{\mu}\mathcal{Z}_{n,\kappa}^{\left(\alpha,\beta\right)}\left(\mu,\xi\right)-\frac{1}{2}\xi\partial_{\xi}\mathcal{Z}_{n,\kappa}^{\left(\alpha,\beta\right)}\left(\mu,\xi\right)\\
=\left(\lambda-2\right)\mathcal{Z}_{n,\kappa}^{\left(\alpha,\beta\right)}\left(\mu,\xi\right)-\kappa\mathcal{Z}_{n,\kappa}^{\left(\alpha,\beta+1\right)}\left(\mu,\xi\right).
\end{multline*}
Hence, after inserting (\ref{eq:GZM16:dZ_cal-dxi}) there results
\begin{multline}
\mu\partial_{\mu}\mathcal{Z}_{n,\kappa}^{\left(\alpha,\beta\right)}\left(\mu,\xi\right)=\left(\lambda-2\right)\mathcal{Z}_{n,\kappa}^{\left(\alpha,\beta\right)}\left(\mu,\xi\right)\\
-\kappa\left(1+\frac{\xi^{2}}{\kappa}\right)\mathcal{Z}_{n,\kappa}^{\left(\alpha,\beta+1\right)}\left(\mu,\xi\right)\\
-\frac{\Gamma\left(\lambda-\nicefrac{1}{2}\right)}{\kappa^{\beta+1}\Gamma\left(\sigma-\nicefrac{3}{2}\right)}\xi\mathcal{H}_{n,\kappa}^{\left(\alpha,\beta+\nicefrac{1}{2}\right)}\left(\mu\right).\label{eq:GZM16:dZ_cal-dmu}
\end{multline}

\begin{widetext}
Finally, the crossed derivative can be obtained from either of the
results above, leading directly to
\begin{multline}
\partial_{\xi,\mu}^{2}\mathcal{Z}_{n,\kappa}^{\left(\alpha,\beta\right)}\left(\mu,\xi\right)=2\frac{\xi}{\mu}\left[\kappa\left(1+\frac{\xi^{2}}{\kappa}\right)\mathcal{Z}_{n,\kappa}^{\left(\alpha,\beta+2\right)}\left(\mu,\xi\right)-\left(\lambda-1\right)\mathcal{Z}_{n,\kappa}^{\left(\alpha,\beta+1\right)}\left(\mu,\xi\right)\right]\\
+\frac{2\Gamma\left(\lambda-\nicefrac{1}{2}\right)\mu^{-1}}{\kappa^{\beta+1}\Gamma\left(\sigma-\nicefrac{3}{2}\right)}\left[\left(\lambda-\frac{1}{2}\right)\left(1+\frac{\xi^{2}}{\kappa}\right)\mathcal{H}_{n,\kappa}^{\left(\alpha,\beta+\nicefrac{3}{2}\right)}\left(\mu\right)-\left(\lambda-\frac{3}{2}\right)\mathcal{H}_{n,\kappa}^{\left(\alpha,\beta+\nicefrac{1}{2}\right)}\left(\mu\right)\right].\label{eq:GZM16:dZ_cal-dxi,mu}
\end{multline}

\end{widetext}

\end{subequations}

\subsubsection{Values at $\boldsymbol{\xi=0}$ or $\boldsymbol{\mu=0}$}

From the definitions (\ref{eq:GZM16:Kappa-PGF}, \ref{eq:GZM16:Z_cal-2_ints},
\ref{eq:GZM16:Y_cal-2_ints}), and (I.11), we obtain the following
limiting expressions,\begin{subequations}\label{eq:GZM16:ZY_cal-xi,mu=00003D0}
\begin{align}
\mathcal{Z}_{n,\kappa}^{\left(\alpha,\beta\right)}\left(0,\xi\right) & =Z_{\kappa}^{\left(\alpha,\beta-1\right)}\left(\xi\right)\delta_{n0}\\
\mathcal{Z}_{n,\kappa}^{\left(\alpha,\beta\right)}\left(\mu,0\right) & =\frac{i\sqrt{\pi}\Gamma\left(\lambda-1\right)}{\kappa^{\beta+1/2}\Gamma\left(\sigma-\nicefrac{3}{2}\right)}\mathcal{H}_{n,\kappa}^{\left(\alpha,\beta\right)}\left(\mu\right)\\
\mathcal{Y}_{n,\kappa}^{\left(\alpha,\beta\right)}\left(\mu,0\right) & =\frac{i\sqrt{\pi}\Gamma\left(\lambda-2\right)}{\kappa^{\beta-1/2}\Gamma\left(\sigma-\nicefrac{3}{2}\right)}\mathcal{H}_{n,\kappa}^{\left(\alpha,\beta-1\right)\prime}\left(\mu\right)\\
\mathcal{Y}_{n,\kappa}^{\left(\alpha,\beta\right)}\left(0,\xi\right) & =-\left(\delta_{n,0}-\frac{1}{2}\delta_{|n|,1}\right)Z_{\kappa}^{\left(\alpha,\beta-3\right)}\left(\xi\right)\\
\partial_{\xi}\mathcal{Z}_{n,\kappa}^{\left(\alpha,\beta\right)}\left(0,\xi\right) & =Z_{\kappa}^{\left(\alpha,\beta-1\right)\prime}\left(\xi\right)\delta_{n0}\\
\partial_{\mu}\mathcal{Z}_{n,\kappa}^{\left(\alpha,\beta\right)}\left(\mu,0\right) & =\frac{i\sqrt{\pi}\Gamma\left(\lambda-1\right)}{\kappa^{\beta+1/2}\Gamma\left(\sigma-\nicefrac{3}{2}\right)}\mathcal{H}_{n,\kappa}^{\left(\alpha,\beta\right)\prime}\left(\mu\right).
\end{align}
\end{subequations}

\subsubsection{Series representations\label{sub:GZM16:ZY_cal-Series-representations}}

In Paper I, we have obtained representations for the functions $\mathcal{Z}_{n,\kappa}^{\left(\alpha,\beta\right)}\left(\mu,\xi\right)$
and $\mathcal{Y}_{n,\kappa}^{\left(\alpha,\beta\right)}\left(\mu,\xi\right)$
in terms of series involving the $\kappa$PGF $\mathcal{H}_{n,\kappa}^{\left(\alpha,\beta\right)}\left(\mu\right)$
and derivatives of the $\kappa$PDF $Z_{\kappa}^{\left(\alpha,\beta\right)}\left(\xi\right)$\@.
These representations are given by Eqs. (I.28a-28b)\@. Subsequent
applications have shown that these expansions start to converge slower
when $\xi_{i}\to-\frac{1}{2}\sqrt{\kappa}$ ($\xi_{i}$: imaginary
part of $\xi$) and may diverge when $\xi_{i}\geqslant-\frac{1}{2}\sqrt{\kappa}$\@.
Consequently, new function representations are necessary, in order
to enlarge the convergence region of the expansions.

In this section, some new expansions for the 2VKPFs are derived. Some
of the obtained expansions are applicable to particular regions of
the functions's domain and some are valid throughout the domain. However,
all representations that have been found have in common that at least
one series expansion is involved, which contains at least one special
function. This is due to the fact that we were not able to factor
the functions in two simpler terms, i.e., $\mathcal{Z}\left(\mu,\xi\right)\neq F_{1}\left(\mu\right)F_{2}\left(\xi\right)$,
for instance. Indeed, we believe that the functions $\mathcal{Z}\left(\mu,\xi\right)$
and $\mathcal{Y}\left(\mu,\xi\right)$ are in fact altogether non-separable.

The transcendental relation between the variables $\mu$ $\left(\sim w_{\perp}\right)$
and $\xi$ $\left(\sim w_{\parallel}\right)$ ultimately stems from
the physical nature of the $\kappa$VDF (\ref{eq:KAP7:f_s})\@. According
to the interpretation of Tsallis's entropic principle, one-particle
distribution functions such as (\ref{eq:KAP7:f_s}) describe the statistical
distribution of particles in a (almost) noncollisional system, but
with a strong correlation between the different degrees of freedom.\cite{LivadiotisMcComas11/11,Livadiotis15/03}
This strong correlation prevents the $\kappa$VDF (\ref{eq:KAP7:f_s})
from being separable in the different velocity components. In contrast,
a physical system in thermal equilibrium has an entropy given by the
Boltzmann-Gibbs statistical mechanics and is characterized by short-range
Coulombian collisions and absence of correlation between the degrees
of freedom. As a consequence, the equilibrium Maxwell-Boltzmann VDF
is completely separable. Therefore, the non-separable nature of the
functions $\mathcal{Z}\left(\mu,\xi\right)$ and $\mathcal{Y}\left(\mu,\xi\right)$
is a mathematical consequence of the strong correlation between different
degrees of freedom of the particles that compose physical systems
statistically described by the $\kappa$VDF.

It is worth mentioning here that the nonadditive statistical mechanics
also admits that particles without correlations may be statistically
described by separable one-particle distribution functions\@.\cite{Livadiotis15/03}
This is the case of the product-bi-kappa (or product-bi-Lorentzian)
VDF,\cite{SummersThorne91/08,Livadiotis15/03} of which the kappa-Maxwellian
distribution\cite{HellbergMace02/05,Cattaert+07/08} is a particular
case. For such distributions, the functions $\mathcal{Z}\left(\mu,\xi\right)$
and $\mathcal{Y}\left(\mu,\xi\right)$ result completely separable
and the mathematical treatment is much simpler. Future works will
also consider this possibility.

The first representation to be derived is a power series in $\xi$,
valid when $\left|\xi\right|<\sqrt{\kappa}$\@. Starting from (\ref{eq:GZM16:Z_cal-1_int-Z}),
we introduce the form (I.15) for the $\kappa$PDF and obtain
\begin{multline*}
\mathcal{Z}_{n,\kappa}^{\left(\alpha,\beta\right)}\left(\mu,\xi\right)=-\frac{4\Gamma\left(\lambda-\nicefrac{1}{2}\right)\xi}{\kappa^{\beta+1}\Gamma\left(\sigma-\nicefrac{3}{2}\right)}\\
\times\int_{0}^{\infty}dx\,\frac{xJ_{n}^{2}\left(\nu x\right)}{\left(1+x^{2}/\kappa\right)^{\lambda-1/2}}\pFq 21\left({1,\lambda-\nicefrac{1}{2}\atop \nicefrac{3}{2}};-\frac{\xi^{2}/\kappa}{1+x^{2}/\kappa}\right)\\
+\frac{2i\pi^{1/2}\Gamma\left(\lambda-1\right)}{\kappa^{\beta+1/2}\Gamma\left(\sigma-\nicefrac{3}{2}\right)}\int_{0}^{\infty}dx\,\frac{xJ_{n}^{2}\left(\nu x\right)}{\left(1+\xi^{2}/\kappa+x^{2}/\kappa\right)^{\lambda-1}}.
\end{multline*}

The second integral can be evaluated. If we initially assume that
$\xi$ is real and define a new integration variable by $x=\sqrt{\psi}t$,
where $\psi=1+\xi^{2}/\kappa$, then we can identify the resulting
integral with (\ref{eq:GZM16:Kappa-PGF}) and write
\begin{equation}
\int_{0}^{\infty}\frac{xJ_{n}^{2}\left(\nu x\right)dx}{\left(1+\xi^{2}/\kappa+x^{2}/\kappa\right)^{\lambda-1}}=\frac{\mathcal{H}_{n,\kappa}^{\left(\alpha,\beta\right)}\left[\mu\left(1+\xi^{2}/\kappa\right)\right]}{2\left(1+\xi^{2}/\kappa\right)^{\lambda-2}}.\label{eq:GZM16:Z_cal-int2}
\end{equation}

Identity (\ref{eq:GZM16:Z_cal-int2}) can be analytically continued
to the complex plane of $\xi$ as long as it stays within the principal
branch of $\mathcal{H}_{n,\kappa}^{\left(\alpha,\beta\right)}\left(z\right)$
(i.e., of the $G$-function)\@. Since the origin is a branch point
of the $G$-function and the infinity is an essential singularity,\cite{Luke75}
the complex-valued $\mathcal{H}$-function in (\ref{eq:GZM16:Z_cal-int2})
has branch cuts along the lines $\left(-i\infty,-i\sqrt{\kappa}\right]$
and $\left[i\sqrt{\kappa},i\infty\right)$\@. Hence, we can employ
result (\ref{eq:GZM16:Z_cal-int2}) when $\left|\xi\right|<\sqrt{\kappa}$\@.

On the other hand, if the Gauss function in the above expression for
$\mathcal{Z}$ is substituted by its power series (I.B4), the series
will also converge if $\left|\xi\right|<\sqrt{\kappa}$, and we are
then allowed to integrate term by term and obtain\begin{subequations}\label{eq:GZM16:Z_cal-series}
\begin{multline}
\mathcal{Z}_{n,\kappa}^{\left(\alpha,\beta\right)}\left(\mu,\xi\right)=-\frac{2\Gamma\left(\lambda-\nicefrac{1}{2}\right)\xi}{\kappa^{\beta+1}\Gamma\left(\sigma-\nicefrac{3}{2}\right)}\\
\times\sum_{k=0}^{\infty}\frac{\left(\lambda-\nicefrac{1}{2}\right)_{k}}{\left(\nicefrac{3}{2}\right)_{k}}\left(-\frac{\xi^{2}}{\kappa}\right)^{k}\mathcal{H}_{n,\kappa}^{\left(\alpha,\beta+k+1/2\right)}\left(\mu\right)\\
+\frac{i\pi^{1/2}\Gamma\left(\lambda-1\right)}{\kappa^{\beta+1/2}\Gamma\left(\sigma-\nicefrac{3}{2}\right)}\frac{\mathcal{H}_{n,\kappa}^{\left(\alpha,\beta\right)}\left[\mu\left(1+\xi^{2}/\kappa\right)\right]}{\left(1+\xi^{2}/\kappa\right)^{\lambda-2}}.\label{eq:GZM16:Z_cal-series-1}
\end{multline}

For the next series expansions, we will consider the $\mathcal{H}$-function
in (\ref{eq:GZM16:Z_cal-1_int-H})\@. Since $1+s^{2}/\kappa\geqslant1$,
we can use the multiplication theorem (\ref{eq:GZM16:G-Mult_theorems})
to write
\begin{multline}
\mathcal{H}_{n,\kappa}^{\left(\alpha,\beta\right)}\left[\mu\left(1+\frac{s^{2}}{\kappa}\right)\right]=\left(1+\frac{s^{2}}{\kappa}\right)^{-\frac{1}{2}}\\
\times\sum_{k=0}^{\infty}\frac{1}{k!}\left(\frac{s^{2}}{\kappa}\right)^{k}\left(1+\frac{s^{2}}{\kappa}\right)^{-k}\widetilde{\mathcal{H}}_{n,k,\kappa}^{\left(\alpha,\beta\right)}\left(\mu\right),\label{eq:GZM16:H-H_associate}
\end{multline}
In this result, the function $\widetilde{\mathcal{H}}_{n,k,\kappa}^{\left(\alpha,\beta\right)}\left(\mu\right)$
is the \emph{associated} \emph{plasma gyroradius function}, defined
by (\ref{eq:GZM16:H-associated}).

In this way, the $\mathcal{Z}$-function can be written in the generic
(and compact) form
\begin{equation}
\mathcal{Z}_{n,\kappa}^{\left(\alpha,\beta\right)}\left(\mu,\xi\right)=\sum_{k=0}^{\infty}\frac{1}{k!}\widetilde{\mathcal{H}}_{n,k,\kappa}^{\left(\alpha,\beta\right)}\left(\mu\right)\widetilde{Z}_{k,\kappa}^{\left(\alpha,\beta\right)}\left(\xi\right),\label{eq:GZM16:Z_cal-HZ_tilde}
\end{equation}
where, accordingly, the function $\widetilde{Z}_{k,\kappa}^{\left(\alpha,\beta\right)}\left(\xi\right)$
is the \emph{associated plasma dispersion function}, defined by (\ref{eq:GZM16:Z-associated}).

Therefore, we can evaluate the function $\mathcal{Z}_{n,\kappa}^{\left(\alpha,\beta\right)}\left(\mu,\xi\right)$
using for $\widetilde{\mathcal{H}}_{n,k,\kappa}^{\left(\alpha,\beta\right)}\left(\mu\right)$
the representations (\ref{eq:GZM16:dH->H_k-RR}) or (\ref{eq:GZM16:H_k-1F2}),
and for $\widetilde{Z}_{k,\kappa}^{\left(\alpha,\beta\right)}\left(\xi\right)$
the representations (\ref{eq:GZM16:Z_tilde-lambda-half-integer}-\ref{eq:GZM16:Z_tilde-Z-k>=00003D1})\@. 

For the $\widetilde{Z}$-function, we can also employ representation
(\ref{eq:eq:GZM16:Z_tilde-2F1}); however, in this case, as was then
mentioned, we also need to include the analytical continuation when
$\xi_{i}=\Im\xi\leqslant0$\@. The necessary expressions can be gleaned
from the discussion concerning the related continuation of Eq. (I.13)\@.
In this process, one would have to include the continuation of the
Gauss function. Alternatively, one can start anew from Eq. (\ref{eq:GZM16:Z_cal-2_ints})
and introduce the adequate continuation for the $s$-integration.
In this way, one would end up with an additional term, which is proportional
to Eq. (\ref{eq:GZM16:Z_cal-int2})\@. Proceeding in this way, the
last series expansion for the $\mathcal{Z}$-function is finally
\begin{multline}
\mathcal{Z}_{n,\kappa}^{\left(\alpha,\beta\right)}\left(\mu,\xi\right)=\sum_{k=0}^{\infty}\frac{1}{k!}\widetilde{\mathcal{H}}_{n,k,\kappa}^{\left(\alpha,\beta\right)}\left(\mu\right)\widetilde{Z}_{(\ref{eq:eq:GZM16:Z_tilde-2F1})}^{\left(\beta\right)}\left(\xi\right)\\
+\frac{2\sqrt{\pi}i\Gamma\left(\lambda-1\right)\Theta\left(-\xi_{i}\right)}{\kappa^{\beta+1/2}\Gamma\left(\sigma-\nicefrac{3}{2}\right)}\left(1+\frac{\xi^{2}}{\kappa}\right)^{-\left(\lambda-2\right)}\\
\times\mathcal{H}_{n,\kappa}^{\left(\alpha,\beta\right)}\left[\mu\left(1+\frac{\xi^{2}}{\kappa}\right)\right],\label{eq:GZM16:Z_cal-HZ_tilde-2F1}
\end{multline}
where we have used the shorthand notation $\widetilde{Z}_{(\ref{eq:eq:GZM16:Z_tilde-2F1})}^{\left(\beta\right)}\left(\xi\right)\equiv\widetilde{Z}_{k,\kappa}^{\left(\alpha,\beta\right)}\left(\xi;\mathrm{Eq.}\ref{eq:eq:GZM16:Z_tilde-2F1}\right)$\@.
We have also employed the Heaviside function $\Theta\left(x\right)=+1$
(if $x\geqslant0$) or $\Theta\left(x\right)=0$ (if $x<0$).

\end{subequations}

The series expansions for the function $\mathcal{Y}_{n,\kappa}^{\left(\alpha,\beta\right)}\left(\mu,\xi\right)$
follow the same methodologies and their derivations will not be repeated.
The only intermediate result shown here is the identity
\begin{multline}
\int_{0}^{\infty}dx\,\frac{x^{3}J_{n-1}\left(\nu x\right)J_{n+1}\left(\nu x\right)}{\left(1+\xi^{2}/\kappa+x^{2}/\kappa\right)^{-\left(\lambda-1\right)}}=\frac{1}{2}\frac{\kappa\mu}{\lambda-2}\\
\times\left(1+\frac{\xi^{2}}{\kappa}\right)^{-\left(\lambda-4\right)}\mathcal{H}_{n,\kappa}^{\left(\alpha,\beta-1\right)\prime}\left[\mu\left(1+\frac{\xi^{2}}{\kappa}\right)\right],\label{eq:GZM16:Y_cal-int2}
\end{multline}
which is derived similarly to Eq. (\ref{eq:GZM16:Z_cal-int2}) and
to which apply the same considerations about the analiticity domain.

Without further ado, the series expansions for $\mathcal{Y}_{n,\kappa}^{\left(\alpha,\beta\right)}\left(\mu,\xi\right)$
are:\begin{subequations}\label{eq:GZM16:Y_cal-series}
\begin{multline}
\mathcal{Y}_{n,\kappa}^{\left(\alpha,\beta\right)}\left(\mu,\xi\right)=-\frac{2\Gamma\left(\lambda-\nicefrac{3}{2}\right)\xi}{\kappa^{\beta}\Gamma\left(\sigma-\nicefrac{3}{2}\right)}\sum_{k=0}^{\infty}\frac{\left(\lambda-\nicefrac{3}{2}\right)_{k}}{\left(\nicefrac{3}{2}\right)_{k}}\\
\times\mathcal{H}_{n,\kappa}^{\left(\alpha,\beta+k-\nicefrac{1}{2}\right)\prime}\left(\mu\right)\left(-\frac{\xi^{2}}{\kappa}\right)^{k}+\frac{i\pi^{1/2}\Gamma\left(\lambda-2\right)}{\kappa^{\beta-1/2}\Gamma\left(\sigma-\nicefrac{3}{2}\right)}\\
\times\frac{\mathcal{H}_{n,\kappa}^{\left(\alpha,\beta-1\right)\prime}\left[\mu\left(1+\xi^{2}/\kappa\right)\right]}{\left(1+\xi^{2}/\kappa\right)^{\lambda-4}},\label{eq:GZM16:Y_cal-series-1}
\end{multline}
valid for $\left|\xi\right|<\sqrt{\kappa}$,
\begin{equation}
\mathcal{Y}_{n,\kappa}^{\left(\alpha,\beta\right)}\left(\mu,\xi\right)=\sum_{k=0}^{\infty}\frac{1}{k!}\widetilde{\mathcal{H}}_{n,k,\kappa}^{\left(\alpha,\beta-1\right)\prime}\left(\mu\right)\widetilde{Z}_{k,\kappa}^{\left(\alpha,\beta-1\right)}\left(\xi\right),\label{eq:GZM16:Y_cal-HZ_tilde}
\end{equation}
valid for any $\xi$, and 
\begin{multline}
\mathcal{Y}_{n,\kappa}^{\left(\alpha,\beta\right)}\left(\mu,\xi\right)=\sum_{k=0}^{\infty}\frac{1}{k!}\widetilde{\mathcal{H}}_{n,k,\kappa}^{\left(\alpha,\beta-1\right)\prime}\left(\mu\right)\widetilde{Z}_{(\ref{eq:eq:GZM16:Z_tilde-2F1})}^{\left(\beta-1\right)}\left(\xi\right)\\
+\frac{2\sqrt{\pi}i\Theta\left(-\xi_{i}\right)\Gamma\left(\lambda-2\right)}{\kappa^{\beta-1/2}\Gamma\left(\sigma-\nicefrac{3}{2}\right)}\left(1+\frac{\xi^{2}}{\kappa}\right)^{-\left(\lambda-4\right)}\\
\times\mathcal{H}_{n,\kappa}^{\left(\alpha,\beta-1\right)\prime}\left[\mu\left(1+\frac{\xi^{2}}{\kappa}\right)\right],\label{eq:GZM16:Y_cal-HZ_tilde-Gauss}
\end{multline}
also valid for any $\xi$.

\end{subequations}

The series expansions and the other properties derived in this section
and in Paper I are sufficient to enable a computational implementation
of the functions $\mathcal{Z}_{n,\kappa}^{\left(\alpha,\beta\right)}\left(\mu,\xi\right)$
and $\mathcal{Y}_{n,\kappa}^{\left(\alpha,\beta\right)}\left(\mu,\xi\right)$,
and hence for the evaluation of the dielectric tensor (\ref{eq:KAP7:DielTensor-1})
for a bi-kappa plasma. 

The numerical evaluation of the series can be substantially accelerated
if one also employs  the recurrence relations (\ref{eq:GZM16:H_k-RR})
and (\ref{eq:GZM16:Z_tilde-RR})\@. However, we must point out that
so far no analysis of the stability of these relations for forward
recursion has been made. It is possible that for a given set of parameters
either or both relations are only stable for backward recursion, and
so different strategies must be implemented.

\subsubsection{Asymptotic expansions}

Here we will derive expressions valid for either $\left|\xi\right|\gg1$
or $\mu\gg1$\@. Starting with $\xi$, the expansion we want to derive
is not the ordinary series representation for $\left|\xi\right|>\sqrt{\kappa}$\@.
Although such a series can be easily obtained from the expressions
already shown, they would be unnecessarily complicated, as it was
hinted by the derivation of the representation (I.16) for the $\kappa$PDF\@.
Instead, we want to derive an expansion valid for $\left|\xi\right|\gg\sqrt{\kappa}$,
convenient for a fluid approximation of the dielectric tensor.

Accordingly, in the $s$-integrals of Eqs. (\ref{eq:GZM16:Z_cal-2_ints})
and (\ref{eq:GZM16:Y_cal-2_ints}) we will approximate
\[
\frac{1}{s-\xi}\stackrel{}{\simeq}-\frac{1}{\xi}\left(1+\frac{s}{\xi}+\frac{s^{2}}{\xi^{2}}+\cdots\right)=-\frac{1}{\xi}\sum_{\ell=0}\frac{s^{\ell}}{\xi^{\ell}},
\]
i.e., we ignore the high-energy particles at the tail of the VDF and
the kinetic effect of the pole at $s=\xi$\@. Notice also that we
have not written the upper limit of the sum above, since such expansion
is only meaningful for a finite number of terms. Inserting this expansion
into the $s$-integrals, all the terms with $\ell$ odd vanish and
the others can be easily evaluated. However, these integrals only
exist if the additional condition $\lambda>k+\nicefrac{3}{2}$ $\left(k=0,1,2,\dots\right)$
is satisfied.

Identifying the remaining $x$-integrals with (\ref{eq:GZM16:Kappa-PGF})
and (\ref{eq:GZM16:Y_cal-int2}), we obtain\begin{subequations}\label{eq:GZM16:ZY_cal-asymp}
\begin{align}
\mathcal{Z}_{n,\kappa}^{\left(\alpha,\beta\right)}\left(\mu,\xi\right) & \simeq-\frac{\pi^{-1/2}\kappa^{-\beta}}{\Gamma\left(\sigma-\nicefrac{3}{2}\right)}\frac{1}{\xi}\sum_{k=0}\Gamma\left(\lambda-k-\frac{3}{2}\right)\nonumber \\
 & \times\Gamma\left(k+\frac{1}{2}\right)\frac{\kappa^{k}}{\xi^{2k}}\mathcal{H}_{n,\kappa}^{\left(\alpha,\beta-k-\nicefrac{1}{2}\right)}\left(\mu\right)\label{eq:GZM16:Z_cal-xi_asympt}\\
\mathcal{Y}_{n,\kappa}^{\left(\alpha,\beta\right)}\left(\mu,\xi\right) & \simeq-\frac{\pi^{-1/2}\kappa^{1-\beta}}{\Gamma\left(\sigma-\nicefrac{3}{2}\right)}\frac{1}{\xi}\sum_{k=0}\Gamma\left(\lambda-k-\frac{5}{2}\right)\nonumber \\
 & \times\Gamma\left(k+\frac{1}{2}\right)\frac{\kappa^{k}}{\xi^{2k}}\mathcal{H}_{n,\kappa}^{\left(\alpha,\beta-k-\nicefrac{3}{2}\right)\prime}\left(\mu\right).\label{eq:GZM16:Y_cal-xi_asympt}
\end{align}

Now, the large gyroradius expansion $\left(\mu\gg1\right)$ is obtained
if we start from (\ref{eq:GZM16:Z_cal-1_int-H}, \ref{eq:GZM16:Y_cal-1_int-H})
and introduce the expansion (\ref{eq:GZM16:H-asymptotic})\@. The
resulting integrals can be identified with the definition of the $\kappa$PDF
in (I.11)\@. Hence, we obtain
\begin{align}
\mathcal{Z}_{n,\kappa}^{\left(\alpha,\beta\right)}\left(\mu,\xi\right)\simeq & \frac{1}{\sqrt{2\pi\mu}}\sum_{k=0}\frac{\left(\nicefrac{1}{2}+n\right)_{k}\left(\nicefrac{1}{2}-n\right)_{k}}{k!\left(2\mu\right)^{k}}\nonumber \\
 & \times Z_{\kappa}^{\left(\alpha,\beta+k-\nicefrac{1}{2}\right)}\left(\xi\right)\label{eq:GZM16:Z_cal-mu_asympt}\\
\mathcal{Y}_{n,\kappa}^{\left(\alpha,\beta\right)}\left(\mu,\xi\right)\simeq & \frac{-1}{\sqrt{2\pi\mu^{3}}}\sum_{k=0}\frac{\left(\nicefrac{1}{2}+n\right)_{k}\left(\nicefrac{1}{2}-n\right)_{k}\left(k+\nicefrac{1}{2}\right)}{k!\left(2\mu\right)^{k}}\nonumber \\
 & \times Z_{\kappa}^{\left(\alpha,\beta+k-\nicefrac{3}{2}\right)}\left(\xi\right).\label{eq:GZM16:Y_cal-mu_asympt}
\end{align}

\end{subequations}

\begin{widetext}

\subsubsection{A closed-form expression for $\boldsymbol{\mathcal{Z}_{n,\kappa}^{\left(\alpha,\beta\right)}\left(\mu,\xi\right)}$}

Since $\mathcal{Z}_{n,\kappa}^{\left(\alpha,\beta\right)}\left(\mu,\xi\right)$
and $\mathcal{Y}_{n,\kappa}^{\left(\alpha,\beta\right)}\left(\mu,\xi\right)$
are non-separable functions of two variables, it is a relevant question
whether they can be represented by some special function discussed
in the literature. Here, we will show for $\mathcal{Z}_{n,\kappa}^{\left(\alpha,\beta\right)}\left(\mu,\xi\right)$
that indeed it can be represented in closed, compact form by the relatively
newly-defined Meijer $G$-function of two variables, introduced in
section \ref{sub:GZM16:G(2)}.

Returning to the definition (\ref{eq:GZM16:Z_cal-2_ints}) and defining
the new integration variables  $x=\sqrt{\kappa u}$ and $s=\sqrt{\kappa v}$,
the double integral can be written as 
\[
I_{2}=\frac{\sqrt{\kappa}}{4}\int_{0}^{\infty}du\,\int_{0}^{\infty}dv\,v^{-1/2}J_{n}^{2}\left(\sqrt{2\kappa\mu u}\right)\frac{\left(1+u+v\right)^{-\left(\lambda-1\right)}}{v-\xi^{2}/\kappa}.
\]
Introducing now the function representations (\ref{eq:GZM16:(1+x)^-rho-G}),
(I.B15c), and (\ref{eq:GZM16:HY92-T13.1.5}), and then expressing
the last in terms of the double Mellin-Barnes integral (\ref{eq:GZM16:G(2)-def}),
one obtains
\begin{multline*}
I_{2}=-\frac{\kappa^{3/2}\xi^{-2}}{4\sqrt{\pi}\Gamma\left(\lambda-1\right)}\frac{1}{\left(2\pi i\right)^{2}}\int_{L_{s}}\int_{L_{t}}dsdt\,\Gamma\left(\lambda-1-s-t\right)\Gamma\left(s\right)\Gamma\left(t\right)\\
\times\left\{ \int_{0}^{\infty}du\,u^{-s}G_{1,3}^{1,1}\left[2\kappa\mu u\left|{\nicefrac{1}{2}\atop n,-n,0}\right.\right]\right\} \left\{ \int_{0}^{\infty}dv\,v^{-t-1/2}G_{1,1}^{1,1}\left[-\frac{\kappa v}{\xi^{2}}\left|{0\atop 0}\right.\right]\right\} ,
\end{multline*}
where we have also interchanged the order of integrations.

The $u$- and $v$-integrations can now be performed by means of the
Mellin transform (\ref{eq:GZM16:G-Mellin_transf}), resulting
\begin{multline}
I_{2}=\frac{\sqrt{\kappa}\left(2\kappa\mu\right)^{-1}}{4\sqrt{\pi}\Gamma\left(\lambda-1\right)}\left(-\frac{\kappa}{\xi^{2}}\right)^{1/2}\frac{1}{\left(2\pi i\right)^{2}}\int_{L_{s}}\int_{L_{t}}dsdt\,\Gamma\left(\lambda-1-s-t\right)\\
\times\frac{\Gamma\left(-\nicefrac{1}{2}+s\right)\Gamma\left(n+1-s\right)}{\Gamma\left(n+s\right)}\Gamma\left(t\right)\Gamma\left(\frac{1}{2}+t\right)\Gamma\left(\frac{1}{2}-t\right)\left[\left(2\kappa\mu\right)^{-1}\right]^{-s}\left(-\frac{\xi^{2}}{\kappa}\right)^{-t}.\label{eq:GZM16:I_2}
\end{multline}
This result can be compared with (\ref{eq:GZM16:G(2)-def}), in which
case we obtain finally
\begin{equation}
\mathcal{Z}_{n,\kappa}^{\left(\alpha,\beta\right)}\left(\mu,\xi\right)=-\frac{\pi^{-1}\kappa^{1-\beta}}{\Gamma\left(\sigma-\nicefrac{3}{2}\right)\xi}G_{1,0:2,1:1,2}^{0,1:1,1:2,1}\left[{\left(2\kappa\mu\right)^{-1}\atop -\xi^{2}/\kappa}\left|{\nicefrac{7}{2}-\lambda:1-n,1+n:1\atop -:\nicefrac{1}{2}:\nicefrac{1}{2},1}\right.\right].\label{eq:GZM16:Z_cal-G(2)}
\end{equation}
The final expression for the $\mathcal{Z}$-function in (\ref{eq:GZM16:Z_cal-G(2)})
was obtained after employing also the translation property (\ref{eq:GZM16:G(2)-translation}).

The Maxwellian limit of (\ref{eq:GZM16:Z_cal-G(2)}) can be obtained.
Expressing again the $G^{(2)}$-function in (\ref{eq:GZM16:Z_cal-G(2)})
in terms of the definition (\ref{eq:GZM16:G(2)-def}), and applying
the limit $\kappa\to\infty$ on the resulting expression, one can
evaluate the limit using Stirling's formula\@.\cite{AskeyRoy-NIST10}
As a result, the $s$- and $t$-integrations factor out, and the remaining
integrals can be identified with $G$-functions from the definition
(I.B10), which in turn can be identified with representations (I.B15d)
and (\ref{eq:GZM16:U-G})\@. After employing properties (I.B11a),
one finally obtains
\[
\lim_{\kappa\to\infty}\mathcal{Z}_{n,\kappa}^{\left(\alpha,\beta\right)}\left(\mu,\xi\right)=e^{-\mu}I_{n}\left(\mu\right)\xi U\left({1\atop \nicefrac{3}{2}};-\xi^{2}\right)=\mathscr{H}_{n}\left(\mu\right)Z\left(\xi\right),
\]
as expected.

Formula (\ref{eq:GZM16:Z_cal-G(2)}) is the more compact representation
of the function $\mathcal{Z}_{n,\kappa}^{\left(\alpha,\beta\right)}\left(\mu,\xi\right)$
that we have obtained. However, despite of being a closed-form for
$\mathcal{Z}$, this representation is not yet very useful, since
there is no known computational implementation that evaluates the
$G^{(2)}$-function, contrary to the one-variable $G$, which is implemented
by some Computer Algebra Software and also by the \texttt{python}
library mpmath\@.\cite{mpmath} Nevertheless, we find it important
to include the derivation of formula (\ref{eq:GZM16:Z_cal-G(2)})
in order to stress the necessity of further development on the numerical
evaluation of these special functions and also to present to the plasma
physics community the techniques involved with Meijer's $G$- and
$G^{(2)}$-functions and Mellin-Barnes integrals in general, since
we believe that as more complex aspects of the physics of plasmas
are considered, such as more general VDFs and dusty plasmas, for instance,
the techniques employed in this work and in Paper I have the potential
to provide mathematical answers to the challenges that will appear.

\end{widetext}

\section{Conclusions\label{sec:Conclusions}}

In this paper we have presented two major developments for the study
of waves with arbitrary frequency and direction of propagation in
anisotropic superthermal plasmas. First, we have derived the dielectric
tensor of a bi-kappa plasma. This tensor will be employed in future
studies concerning wave propagation and amplification/damping in anisotropic
superthermal plasmas.

The tensor components were written in terms of the kappa plasma special
functions, which must be numerically evaluated for practical applications.
To this end, we have derived in the present paper (and in Paper I)
several mathematical properties and representations for these functions.
With the development presented here and in Paper I, we believe that
all the necessary framework for a systematic study of electromagnetic/electrostatic
waves propagating at arbitrary angles in a bi-kappa plasma has been
obtained. In future studies we will apply this formalism to specific
problems concerning temperature-driven-instabilities in kappa plasmas.
\begin{acknowledgments}
The authors acknowledge support provided by Conselho Nacional de Desenvolvimento
Científico e Tecnológico (CNPq), grants No. 304461/2012-1, 478728/2012-3
and 304363/2014-6.
\end{acknowledgments}

\appendix

\section{Derivation of the susceptibility tensor\label{sec:GZM16:SuscTensor-derivation}}

The derivation of $\chi_{ij}^{(s)}$ for a bi-kappa plasma (or for
any VDF, for that matter) is simplified if one observes that all tensor
components have common factors. First, inserting the function (\ref{eq:KAP7:f_s})
into the tensor (\ref{eq:SuscTensor}), all components contain the
derivatives $Lf_{s}$ and $\mathcal{L}f_{s}$\@. Using these derivatives,
one can proceed with the evaluation of the integrals. Using a cylindrical
coordinate system and defining the nondimensional integration variables
$t=v_{\perp}/w_{\perp s}$ and $u=v_{\parallel}/w_{\parallel s}$,
one obtains, after some straightforward algebra, the unified form
\begin{multline*}
\chi_{ij}^{(s)}=2\frac{\omega_{ps}^{2}}{\omega^{2}}\frac{\sigma_{s}g\left(\kappa_{s},\alpha_{s}\right)}{\pi^{1/2}\kappa_{s}}\sum_{n\to-\infty}^{\infty}\\
\times\int_{0}^{\infty}dt\,\int_{-\infty}^{\infty}du\,I_{ij,n}^{(s)}\left(1+\frac{u^{2}}{\kappa_{s}}+\frac{t^{2}}{\kappa_{s}}\right)^{-\sigma_{s}-1},
\end{multline*}
where
\begin{align*}
I_{ij,n}^{(s)} & =\left(\xi_{0s}-A_{s}u\right)J_{ij,n}^{(s)}, & I_{iz,n}^{(s)} & =\left(\xi_{0s}-A_{s}\xi_{ns}\right)K_{ij,n}^{(s)},
\end{align*}
 
\begin{align*}
J_{xx,n}^{(s)} & =\frac{n^{2}}{\mu_{s}}\frac{tJ_{n}^{2}\left(\nu_{s}t\right)}{u-\xi_{ns}}\\
J_{xy,n}^{(s)} & =\sqrt{2}i\frac{n}{\mu_{s}}\frac{t^{2}J_{n}\left(\nu_{s}t\right)J_{n}^{\prime}\left(\nu_{s}t\right)}{u-\xi_{ns}}\\
J_{yy,s}^{(s)} & =2\frac{\frac{n^{2}}{2\mu_{s}}tJ_{n}^{2}\left(\nu_{s}t\right)-t^{3}J_{n-1}\left(\nu_{s}t\right)J_{n+1}\left(\nu_{s}t\right)}{u-\xi_{ns}}\\
K_{xz,n}^{(s)} & =\sqrt{2}\frac{w_{\parallel s}}{w_{\perp s}}\frac{n}{\mu_{s}}\frac{tuJ_{n}^{2}\left(\nu_{s}t\right)}{u-\xi_{ns}}\\
K_{yz,n}^{(s)} & =-2i\frac{w_{\parallel s}}{w_{\perp s}}\frac{t^{2}uJ_{n}\left(\nu_{s}t\right)J_{n}^{\prime}\left(\nu_{s}t\right)}{u-\xi_{ns}}\\
K_{zz,n}^{(s)} & =2\frac{w_{\parallel s}^{2}}{w_{\perp s}^{2}}\xi_{ns}\frac{tuJ_{n}^{2}\left(\nu_{s}t\right)}{u-\xi_{ns}},
\end{align*}
with $g\left(\kappa_{s},\alpha_{s}\right)=\kappa_{s}^{-3/2}\Gamma\left(\sigma_{s}\right)/\Gamma\left(\sigma_{s}-\nicefrac{3}{2}\right)$,
$\nu_{s}=k_{\perp}w_{\perp s}/\Omega_{s}$, and where the anisotropy
parameter $A_{s}=1-w_{\perp s}^{2}/w_{\parallel s}^{2}$ appears for
the first time\@. These results were obtained using the identity
(\ref{eq:GZM16:BesselJ-sum}) and the recurrence relations of the
Bessel functions.

The remaining integrals in the $J$s and $K$s can now be identified
with the definitions of the two-variables kappa plasma functions $\mathcal{Z}_{n,\kappa}^{\left(\alpha,\beta\right)}\left(\mu,\xi\right)$
and $\mathcal{Y}_{n,\kappa}^{\left(\alpha,\beta\right)}\left(\mu,\xi\right)$
and their derivatives, given by Eqs. (\ref{eq:GZM16:ZY_cal}) and
(\ref{eq:GZM16:Z_cal-ders})\@. In this way, one arrives at the final
expressions shown in Eqs. (\ref{eq:ST-BK-1}-\ref{eq:ST-BK-6}).

The Maxwellian limit of the partial susceptibility tensor is obtained
by the process $\kappa_{s}\to\infty$\@. Upon applying this limit,
one must replace $w_{\parallel(\perp)}\to v_{T\parallel(\perp)}=\sqrt{2T_{\parallel(\perp)}/m}$
and the kappa plasma functions are replaced by their limiting representations
(\ref{eq:GZM16:ZY_cal-Maxwellian})\@. In this way, one arrives at\begin{subequations}\label{eq:SuscTensor_Bi-Maxwellian}
\begin{align}
\chi_{xx}^{(s)}= & \frac{\omega_{ps}^{2}}{\omega^{2}}\sum_{n\to-\infty}^{\infty}\frac{n^{2}}{\mu_{s}}\mathscr{H}_{n}\left(\mu_{s}\right)\left[\xi_{0s}Z\left(\xi_{ns}\right)+\frac{1}{2}A_{s}Z^{\prime}\left(\xi_{ns}\right)\right]\\
\chi_{xy}^{(s)}= & \,i\frac{\omega_{ps}^{2}}{\omega^{2}}\sum_{n\to-\infty}^{\infty}n\mathscr{H}_{n}^{\prime}\left(\mu_{s}\right)\left[\xi_{0s}Z\left(\xi_{ns}\right)+\frac{1}{2}A_{s}Z^{\prime}\left(\xi_{ns}\right)\right]\\
\chi_{xz}^{(s)}= & -\frac{\omega_{ps}^{2}}{\omega^{2}}\frac{v_{T\parallel s}}{v_{T\perp s}}\sum_{n\to-\infty}^{\infty}\frac{n\Omega_{s}}{k_{\perp}v_{T\perp s}}\left(\xi_{0s}-A_{s}\xi_{ns}\right)\nonumber \\
 & \times\mathscr{H}_{n}\left(\mu_{s}\right)Z^{\prime}\left(\xi_{ns}\right)\\
\chi_{yy}^{(s)}= & \frac{\omega_{ps}^{2}}{\omega^{2}}\sum_{n\to-\infty}^{\infty}\left[\frac{n^{2}}{\mu_{s}}\mathscr{H}_{n}\left(\mu_{s}\right)-2\mu_{s}\mathscr{H}_{n}^{\prime}\left(\mu_{s}\right)\right]\nonumber \\
 & \times\left[\xi_{0s}Z\left(\xi_{ns}\right)+\frac{1}{2}A_{s}Z^{\prime}\left(\xi_{ns}\right)\right]\\
\chi_{yz}^{(s)}= & \,i\frac{\omega_{ps}^{2}}{\omega^{2}}\frac{v_{T\parallel s}}{v_{T\perp s}}\frac{k_{\perp}v_{T\perp s}}{2\Omega_{s}}\sum_{n\to-\infty}^{\infty}\left(\xi_{0s}-A_{s}\xi_{ns}\right)\nonumber \\
 & \times\mathscr{H}_{n}^{\prime}\left(\mu_{s}\right)Z^{\prime}\left(\xi_{ns}\right)\\
\chi_{zz}^{(s)}= & -\frac{\omega_{ps}^{2}}{\omega^{2}}\frac{v_{T\parallel s}^{2}}{v_{T\perp s}^{2}}\sum_{n\to-\infty}^{\infty}\left(\xi_{0s}-A_{s}\xi_{ns}\right)\nonumber \\
 & \times\xi_{ns}\mathscr{H}_{n}\left(\mu_{s}\right)Z^{\prime}\left(\xi_{ns}\right).
\end{align}
\end{subequations}These results agree with expressions that can be
found in the literature. See, e.g., Eq. (20) of Ref. \onlinecite{Cattaert+07/08}.

\section{The one- and two-variables Meijer $\boldsymbol{G}$-functions\label{sec:GZM16:G-functions}}

\subsection{The $\boldsymbol{G}$-function}

The definition and some properties of the $G$-function are given
in Sec. B.2 of Paper I\@. All identities shown there and in the following
can be found in Refs. \onlinecite{Luke75,Prudnikov90v3}, except when
explicitly mentioned.

\subsubsection{Derivatives}

We have\begin{subequations}\label{eq:GZM16:G-derivatives}
\begin{equation}
\frac{d^{k}}{dz^{k}}G_{p,q}^{m,n}\left[z\left|{\left(a_{p}\right)\atop \left(b_{q}\right)}\right.\right]=\left(-z\right)^{-k}G_{p+1,q+1}^{m+1,n}\left[z\left|{\left(a_{p}\right),0\atop k,\left(b_{q}\right)}\right.\right].\label{eq:der1}
\end{equation}

\begin{widetext}
We will now derive a formula that is not usually found in the literature.
If $n\geqslant1$, we can employ the definition of the $G$-function
in terms of a Mellin-Barnes integral, given by (I.B10), and evaluate,
for $k=0,1,2,\dots$, 
\[
\frac{d^{k}}{dz^{k}}\left\{ z^{k-a_{1}}G_{p,q}^{m,n}\left[z\left|{a_{1},\dots,a_{p}\atop \left(b_{q}\right)}\right.\right]\right\} =\frac{1}{2\pi i}\int_{L}\frac{\prod_{j=1}^{m}\Gamma\left(b_{j}-s\right)\prod_{j=1}^{n}\Gamma\left(1-a_{j}+s\right)}{\prod_{j=m+1}^{q}\Gamma\left(1-b_{j}+s\right)\prod_{j=n+1}^{p}\Gamma\left(a_{j}-s\right)}\frac{\Gamma\left(1-a_{1}+k+s\right)}{\Gamma\left(1-a_{1}+s\right)}z^{-a_{1}+s}ds,
\]
since $D^{m}z^{\gamma}=\Gamma\left(\gamma+1\right)z^{\gamma-m}/\Gamma\left(\gamma+1-m\right)$\@.
Consequently, we obtain the differentiation formula
\begin{equation}
\frac{d^{k}}{dz^{k}}\left\{ z^{k-a_{1}}G_{p,q}^{m,n}\left[z\left|{a_{1},\dots,a_{p}\atop \left(b_{q}\right)}\right.\right]\right\} =z^{-a_{1}}G_{p,q}^{m,n}\left[z\left|{a_{1}-k,\dots,a_{p}\atop \left(b_{q}\right)}\right.\right]\;\left(n\geqslant1\right).\label{eq:der2}
\end{equation}

\end{widetext}

\end{subequations}

\subsubsection{Multiplication theorems}

If $\Re w>\nicefrac{1}{2}$ and $n>0$, 
\begin{multline}
G_{p,q}^{m,n}\left[zw\left|{\left(a_{p}\right)\atop \left(b_{q}\right)}\right.\right]=w^{a_{1}-1}\\
\times\sum_{k=0}^{\infty}\frac{\left(1-1/w\right)^{k}}{k!}G_{p,q}^{m,n}\left[z\left|{a_{1}-k,a_{2},\dots,a_{p}\atop \left(b_{q}\right)}\right.\right].\label{eq:GZM16:G-Mult_theorems}
\end{multline}

\subsubsection{Mellin transform}

The Mellin transform of the $G$-function is
\begin{multline}
\int_{0}^{\infty}y^{s-1}G_{p,q}^{m,n}\left[\eta y\left|{\left(a_{p}\right)\atop \left(b_{q}\right)}\right.\right]dy\\
=\frac{\prod_{j=1}^{m}\Gamma\left(b_{j}+s\right)\prod_{j=1}^{n}\Gamma\left(1-a_{j}-s\right)\eta^{-s}}{\prod_{j=m+1}^{q}\Gamma\left(1-b_{j}-s\right)\prod_{j=n+1}^{p}\Gamma\left(a_{j}+s\right)}.\label{eq:GZM16:G-Mellin_transf}
\end{multline}

\subsubsection{Function representations}

A short list of function representations is:\begin{subequations}\label{eq:GZM16:Functions-G}
\begin{align}
\left(1+x\right)^{-\rho} & =\frac{1}{\Gamma\left(\rho\right)}G_{1,1}^{1,1}\left[x\left|{1-\rho\atop 0}\right.\right]\label{eq:GZM16:(1+x)^-rho-G}\\
\frac{I_{\mu}\left(\sqrt{z}\right)K_{\nu}\left(\sqrt{z}\right)}{\left(2\sqrt{\pi}\right)^{-1}} & =G_{2,4}^{2,2}\left[z\left|{0,\nicefrac{1}{2}\atop \frac{\mu+\nu}{2},\frac{\mu-\nu}{2},-\frac{\mu-\nu}{2},-\frac{\mu+\nu}{2}}\right.\right]\label{eq:GZM16:I_muK_nu-G}\\
\frac{\Gamma\left(a\right)U\left({a\atop b};z\right)}{\left[\Gamma\left(a-b+1\right)\right]^{-1}} & =G_{1,2}^{2,1}\left[z\left|{1-a\atop 0,1-b}\right.\right].\label{eq:GZM16:U-G}
\end{align}
\end{subequations}

\begin{widetext}

\subsection{The two-variables Meijer function\label{sub:GZM16:G(2)}}

The logical extension of Meijer's $G$-function for two variables
was first proposed by Agarwal\cite{Agarwal65} in 1965\@. Subsequent
publications proposed slightly different definitions for the same
extension\@.\cite{Sharma65b,Verma66/07,HaiYakubovich92} In this
work, we will adopt the definition by Hai and Yakubovich (Eq. 13.1
of Ref. \onlinecite{HaiYakubovich92}),
\begin{equation}
G_{p_{1},q_{1}:p_{2},q_{2}:p_{3},q_{3}}^{m_{1},n_{1}:m_{2},n_{2}:m_{3},n_{3}}\left[{x\atop y}\left|{\left(a_{p_{1}}^{\left(1\right)}\right):\left(a_{p_{2}}^{\left(2\right)}\right):\left(a_{p_{3}}^{\left(3\right)}\right)\atop \left(b_{q_{1}}^{\left(1\right)}\right):\left(b_{q_{2}}^{\left(2\right)}\right):\left(b_{q_{3}}^{\left(3\right)}\right)}\right.\right]=\frac{1}{\left(2\pi i\right)^{2}}\int_{L_{s}}\int_{L_{t}}\Psi_{1}\left(s+t\right)\Psi_{2}\left(s\right)\Psi_{3}\left(t\right)x^{-s}y^{-t}dsdt,\label{eq:GZM16:G(2)-def}
\end{equation}
where, for $k=1,2,3$, 
\[
\Psi_{k}\left(r\right)=\frac{\prod_{j=1}^{m_{k}}\Gamma\left(b_{j}^{\left(k\right)}+r\right)\prod_{j=1}^{n_{k}}\Gamma\left(1-a_{j}^{\left(k\right)}-r\right)}{\prod_{j=n_{k}+1}^{p_{k}}\Gamma\left(a_{j}^{\left(k\right)}+r\right)\prod_{j=m_{k}+1}^{q_{k}}\Gamma\left(1-b_{j}^{\left(k\right)}-r\right)}.
\]
The Reader is referred to Sec. II.13 of Ref. \onlinecite{HaiYakubovich92}
for explanation on the notation and a discussion on the general conditions
on the validity of (\ref{eq:GZM16:G(2)-def})\@. Whenever convenient
and unambiguous, we will refer to the two-variables Meijer function
as the $G^{(2)}$-function.

We list some elementary properties of the $G^{(2)}$-function, some
of which are employed in this work. The symmetry property
\begin{equation}
G_{p_{1},q_{1}:p_{2},q_{2}:p_{3},q_{3}}^{m_{1},n_{1}:m_{2},n_{2}:m_{3},n_{3}}\left[{x\atop y}\left|{\left(a_{p_{1}}^{\left(1\right)}\right):\left(a_{p_{2}}^{\left(2\right)}\right):\left(a_{p_{3}}^{\left(3\right)}\right)\atop \left(b_{q_{1}}^{\left(1\right)}\right):\left(b_{q_{2}}^{\left(2\right)}\right):\left(b_{q_{3}}^{\left(3\right)}\right)}\right.\right]=G_{q_{1},p_{1}:q_{2},p_{2}:q_{3},p_{3}}^{n_{1},m_{1}:n_{2},m_{2}:n_{3},m_{3}}\left[{x^{-1}\atop y^{-1}}\left|{1-\left(b_{q_{1}}^{\left(1\right)}\right):1-\left(b_{q_{2}}^{\left(2\right)}\right):1-\left(b_{q_{3}}^{\left(3\right)}\right)\atop 1-\left(a_{p_{1}}^{\left(1\right)}\right):1-\left(a_{p_{2}}^{\left(2\right)}\right):1-\left(a_{p_{3}}^{\left(3\right)}\right)}\right.\right],\label{eq:GZM16:G(2)-symmetry}
\end{equation}
and the translation property
\begin{multline}
x^{\alpha}y^{\beta}G_{p_{1},q_{1}:p_{2},q_{2}:p_{3},q_{3}}^{m_{1},n_{1}:m_{2},n_{2}:m_{3},n_{3}}\left[{x\atop y}\left|{\left(a_{p_{1}}^{\left(1\right)}\right):\left(a_{p_{2}}^{\left(2\right)}\right):\left(a_{p_{3}}^{\left(3\right)}\right)\atop \left(b_{q_{1}}^{\left(1\right)}\right):\left(b_{q_{2}}^{\left(2\right)}\right):\left(b_{q_{3}}^{\left(3\right)}\right)}\right.\right]\\
=G_{p_{1},q_{1}:p_{2},q_{2}:p_{3},q_{3}}^{m_{1},n_{1}:m_{2},n_{2}:m_{3},n_{3}}\left[{x\atop y}\left|{\left(a_{p_{1}}^{\left(1\right)}\right)+\alpha+\beta:\left(a_{p_{2}}^{\left(2\right)}\right)+\alpha:\left(a_{p_{3}}^{\left(3\right)}\right)+\beta\atop \left(b_{q_{1}}^{\left(1\right)}\right)+\alpha+\beta:\left(b_{q_{2}}^{\left(2\right)}\right)+\alpha:\left(b_{q_{3}}^{\left(3\right)}\right)+\beta}\right.\right].\label{eq:GZM16:G(2)-translation}
\end{multline}

A product of two $G$-functions can be written as a single $G^{(2)}$-function
as
\begin{equation}
G_{0,0:p_{2},q_{2}:p_{3},q_{3}}^{0,0:m_{2},n_{2}:m_{3},n_{3}}\left[{x\atop y}\left|{-:\left(a_{p_{2}}^{\left(2\right)}\right):\left(a_{p_{3}}^{\left(3\right)}\right)\atop -:\left(b_{q_{2}}^{\left(2\right)}\right):\left(b_{q_{3}}^{\left(3\right)}\right)}\right.\right]=G_{p_{2},q_{2}}^{m_{2},n_{2}}\left[x\left|{\left(a_{p_{2}}^{\left(2\right)}\right)\atop \left(b_{q_{2}}^{\left(2\right)}\right)}\right.\right]G_{p_{3},q_{3}}^{m_{3},n_{3}}\left[y\left|{\left(a_{p_{3}}^{\left(3\right)}\right)\atop \left(b_{q_{3}}^{\left(3\right)}\right)}\right.\right].\label{eq:GZM16:G(2)-GG}
\end{equation}

We will also use the function representation
\begin{equation}
\left(1+x+y\right)^{-\alpha}=\frac{1}{\Gamma\left(\alpha\right)}G_{0,1:1,0:1,0}^{1,0:0,1:0,1}\left[{x\atop y}\left|{-:1:1\atop \alpha:-:-}\right.\right].\label{eq:GZM16:HY92-T13.1.5}
\end{equation}

Properties (\ref{eq:GZM16:G(2)-symmetry})-(\ref{eq:GZM16:G(2)-GG})
can be inferred from the definition (\ref{eq:GZM16:G(2)-def})\@.
The identity (\ref{eq:GZM16:HY92-T13.1.5}) is given in Sec. II.13
of Ref. \onlinecite{HaiYakubovich92}.

\end{widetext}

\bibliography{Gaelzer+16a.bib}

\begin{thebibliography}{43}%
\makeatletter
\providecommand \@ifxundefined [1]{%
 \@ifx{#1\undefined}
}%
\providecommand \@ifnum [1]{%
 \ifnum #1\expandafter \@firstoftwo
 \else \expandafter \@secondoftwo
 \fi
}%
\providecommand \@ifx [1]{%
 \ifx #1\expandafter \@firstoftwo
 \else \expandafter \@secondoftwo
 \fi
}%
\providecommand \natexlab [1]{#1}%
\providecommand \enquote  [1]{``#1''}%
\providecommand \bibnamefont  [1]{#1}%
\providecommand \bibfnamefont [1]{#1}%
\providecommand \citenamefont [1]{#1}%
\providecommand \href@noop [0]{\@secondoftwo}%
\providecommand \href [0]{\begingroup \@sanitize@url \@href}%
\providecommand \@href[1]{\@@startlink{#1}\@@href}%
\providecommand \@@href[1]{\endgroup#1\@@endlink}%
\providecommand \@sanitize@url [0]{\catcode `\\12\catcode `\$12\catcode
  `\&12\catcode `\#12\catcode `\^12\catcode `\_12\catcode `\%12\relax}%
\providecommand \@@startlink[1]{}%
\providecommand \@@endlink[0]{}%
\providecommand \url  [0]{\begingroup\@sanitize@url \@url }%
\providecommand \@url [1]{\endgroup\@href {#1}{\urlprefix }}%
\providecommand \urlprefix  [0]{URL }%
\providecommand \Eprint [0]{\href }%
\providecommand \doibase [0]{http://dx.doi.org/}%
\providecommand \selectlanguage [0]{\@gobble}%
\providecommand \bibinfo  [0]{\@secondoftwo}%
\providecommand \bibfield  [0]{\@secondoftwo}%
\providecommand \translation [1]{[#1]}%
\providecommand \BibitemOpen [0]{}%
\providecommand \bibitemStop [0]{}%
\providecommand \bibitemNoStop [0]{.\EOS\space}%
\providecommand \EOS [0]{\spacefactor3000\relax}%
\providecommand \BibitemShut  [1]{\csname bibitem#1\endcsname}%
\let\auto@bib@innerbib\@empty
\bibitem [{\citenamefont {Livadiotis}\ and\ \citenamefont
  {McComas}(2011)}]{LivadiotisMcComas11/11}%
  \BibitemOpen
  \bibfield  {author} {\bibinfo {author} {\bibfnamefont {G.}~\bibnamefont
  {Livadiotis}}\ and\ \bibinfo {author} {\bibfnamefont {D.~J.}\ \bibnamefont
  {McComas}},\ }\href {\doibase 10.1088/0004-637X/741/2/88} {\bibfield
  {journal} {\bibinfo  {journal} {Astrophys. J.}\ }\textbf {\bibinfo {volume}
  {741}},\ \bibinfo {pages} {88} (\bibinfo {year} {2011})}\BibitemShut
  {NoStop}%
\bibitem [{\citenamefont {Livadiotis}\ and\ \citenamefont
  {McComas}(2009)}]{LivadiotisMcComas09/11}%
  \BibitemOpen
  \bibfield  {author} {\bibinfo {author} {\bibfnamefont {G.}~\bibnamefont
  {Livadiotis}}\ and\ \bibinfo {author} {\bibfnamefont {D.~J.}\ \bibnamefont
  {McComas}},\ }\href {\doibase 10.1029/2009JA014352} {\bibfield  {journal}
  {\bibinfo  {journal} {J. Geophys. Res.}\ }\textbf {\bibinfo {volume} {114}}
  (\bibinfo {year} {2009}),\ 10.1029/2009JA014352}\BibitemShut {NoStop}%
\bibitem [{\citenamefont {Livadiotis}\ and\ \citenamefont
  {McComas}(2013)}]{LivadiotisMccomas13/05}%
  \BibitemOpen
  \bibfield  {author} {\bibinfo {author} {\bibfnamefont {G.}~\bibnamefont
  {Livadiotis}}\ and\ \bibinfo {author} {\bibfnamefont {D.~J.}\ \bibnamefont
  {McComas}},\ }\href {\doibase 10.1007/s11214-013-9982-9} {\bibfield
  {journal} {\bibinfo  {journal} {Space Sci. Rev.}\ }\textbf {\bibinfo {volume}
  {175}},\ \bibinfo {pages} {183–214} (\bibinfo {year} {2013})}\BibitemShut
  {NoStop}%
\bibitem [{\citenamefont {Livadiotis}(2015)}]{Livadiotis15/03}%
  \BibitemOpen
  \bibfield  {author} {\bibinfo {author} {\bibfnamefont {G.}~\bibnamefont
  {Livadiotis}},\ }\href {\doibase 10.1002/2014JA020825} {\bibfield  {journal}
  {\bibinfo  {journal} {J. Geophys. Res.}\ }\textbf {\bibinfo {volume} {120}}
  (\bibinfo {year} {2015}),\ 10.1002/2014JA020825}\BibitemShut {NoStop}%
\bibitem [{\citenamefont {Gaelzer}\ and\ \citenamefont
  {Ziebell}(2014)}]{GaelzerZiebell14/12}%
  \BibitemOpen
  \bibfield  {author} {\bibinfo {author} {\bibfnamefont {R.}~\bibnamefont
  {Gaelzer}}\ and\ \bibinfo {author} {\bibfnamefont {L.~F.}\ \bibnamefont
  {Ziebell}},\ }\href {\doibase 10.1002/2014JA020667} {\bibfield  {journal}
  {\bibinfo  {journal} {J. Geophys. Res.}\ }\textbf {\bibinfo {volume} {119}}
  (\bibinfo {year} {2014}),\ 10.1002/2014JA020667}\BibitemShut {NoStop}%
\bibitem [{\citenamefont {Gaelzer}\ and\ \citenamefont
  {Ziebell}(2016)}]{GaelzerZiebell16/02}%
  \BibitemOpen
  \bibfield  {author} {\bibinfo {author} {\bibfnamefont {R.}~\bibnamefont
  {Gaelzer}}\ and\ \bibinfo {author} {\bibfnamefont {L.~F.}\ \bibnamefont
  {Ziebell}},\ }\href {\doibase 10.1063/1.4941260} {\bibfield  {journal}
  {\bibinfo  {journal} {Phys. Plasmas}\ }\textbf {\bibinfo {volume} {23}},\
  \bibinfo {eid} {022110} (\bibinfo {year} {2016}),\
  10.1063/1.4941260}\BibitemShut {NoStop}%
\bibitem [{\citenamefont {Summers}\ and\ \citenamefont
  {Thorne}(1991)}]{SummersThorne91/08}%
  \BibitemOpen
  \bibfield  {author} {\bibinfo {author} {\bibfnamefont {D.}~\bibnamefont
  {Summers}}\ and\ \bibinfo {author} {\bibfnamefont {R.~M.}\ \bibnamefont
  {Thorne}},\ }\href {\doibase 10.1063/1.859653} {\bibfield  {journal}
  {\bibinfo  {journal} {Phys. Fluids B}\ }\textbf {\bibinfo {volume} {3}},\
  \bibinfo {pages} {1835} (\bibinfo {year} {1991})}\BibitemShut {NoStop}%
\bibitem [{\citenamefont {Yoon}, \citenamefont {Wu},\ and\ \citenamefont {{de
  Assis}}(1993)}]{YoonWuAssis93/07}%
  \BibitemOpen
  \bibfield  {author} {\bibinfo {author} {\bibfnamefont {P.~H.}\ \bibnamefont
  {Yoon}}, \bibinfo {author} {\bibfnamefont {C.~S.}\ \bibnamefont {Wu}}, \ and\
  \bibinfo {author} {\bibfnamefont {A.~S.}\ \bibnamefont {{de Assis}}},\ }\href
  {\doibase 10.1063/1.860785} {\bibfield  {journal} {\bibinfo  {journal} {Phys.
  Fluids B}\ }\textbf {\bibinfo {volume} {5}},\ \bibinfo {pages} {1971}
  (\bibinfo {year} {1993})}\BibitemShut {NoStop}%
\bibitem [{\citenamefont {Hellinger}\ and\ \citenamefont
  {Matsumoto}(2000)}]{HellingerMatsumoto00/05}%
  \BibitemOpen
  \bibfield  {author} {\bibinfo {author} {\bibfnamefont {P.}~\bibnamefont
  {Hellinger}}\ and\ \bibinfo {author} {\bibfnamefont {H.}~\bibnamefont
  {Matsumoto}},\ }\href {\doibase 10.1029/1999JA000297} {\bibfield  {journal}
  {\bibinfo  {journal} {J. Geophys. Res.}\ }\textbf {\bibinfo {volume} {105}},\
  \bibinfo {pages} {10519} (\bibinfo {year} {2000})}\BibitemShut {NoStop}%
\bibitem [{\citenamefont {Bashir}\ \emph {et~al.}(2010)\citenamefont {Bashir},
  \citenamefont {Iqbal}, \citenamefont {Aslam},\ and\ \citenamefont
  {Murtaza}}]{Bashir+10/10}%
  \BibitemOpen
  \bibfield  {author} {\bibinfo {author} {\bibfnamefont {M.~F.}\ \bibnamefont
  {Bashir}}, \bibinfo {author} {\bibfnamefont {Z.}~\bibnamefont {Iqbal}},
  \bibinfo {author} {\bibfnamefont {I.}~\bibnamefont {Aslam}}, \ and\ \bibinfo
  {author} {\bibfnamefont {G.}~\bibnamefont {Murtaza}},\ }\href {\doibase
  10.1063/1.3499389} {\bibfield  {journal} {\bibinfo  {journal} {Phys.
  Plasmas}\ }\textbf {\bibinfo {volume} {17}},\ \bibinfo {eid} {102112}
  (\bibinfo {year} {2010}),\ 10.1063/1.3499389}\BibitemShut {NoStop}%
\bibitem [{\citenamefont {Chen}\ and\ \citenamefont {Wu}(2010)}]{ChenWu10/06}%
  \BibitemOpen
  \bibfield  {author} {\bibinfo {author} {\bibfnamefont {L.}~\bibnamefont
  {Chen}}\ and\ \bibinfo {author} {\bibfnamefont {D.~J.}\ \bibnamefont {Wu}},\
  }\href {\doibase 10.1063/1.3439680} {\bibfield  {journal} {\bibinfo
  {journal} {Phys. Plasmas}\ }\textbf {\bibinfo {volume} {17}},\ \bibinfo {eid}
  {062107} (\bibinfo {year} {2010})}\BibitemShut {NoStop}%
\bibitem [{\citenamefont {Chen}, \citenamefont {Wu},\ and\ \citenamefont
  {Huang}(2013)}]{Chen+13/06}%
  \BibitemOpen
  \bibfield  {author} {\bibinfo {author} {\bibfnamefont {L.}~\bibnamefont
  {Chen}}, \bibinfo {author} {\bibfnamefont {D.~J.}\ \bibnamefont {Wu}}, \ and\
  \bibinfo {author} {\bibfnamefont {J.}~\bibnamefont {Huang}},\ }\href
  {\doibase 10.1002/jgra.50332} {\bibfield  {journal} {\bibinfo  {journal} {J.
  Geophys. Res.}\ }\textbf {\bibinfo {volume} {118}},\ \bibinfo {pages}
  {2951–2957} (\bibinfo {year} {2013})}\BibitemShut {NoStop}%
\bibitem [{\citenamefont {Shukla}\ \emph {et~al.}(2008)\citenamefont {Shukla},
  \citenamefont {Mishra}, \citenamefont {Varma},\ and\ \citenamefont
  {Tiwari}}]{Shukla+08/02}%
  \BibitemOpen
  \bibfield  {author} {\bibinfo {author} {\bibfnamefont {N.}~\bibnamefont
  {Shukla}}, \bibinfo {author} {\bibfnamefont {R.}~\bibnamefont {Mishra}},
  \bibinfo {author} {\bibfnamefont {P.}~\bibnamefont {Varma}}, \ and\ \bibinfo
  {author} {\bibfnamefont {M.~S.}\ \bibnamefont {Tiwari}},\ }\href {\doibase
  10.1088/0741-3335} {\bibfield  {journal} {\bibinfo  {journal} {Plasma Phys.
  Cont. Fusion}\ }\textbf {\bibinfo {volume} {50}},\ \bibinfo {eid} {025001}
  (\bibinfo {year} {2008})}\BibitemShut {NoStop}%
\bibitem [{\citenamefont {Naim}, \citenamefont {Bashir},\ and\ \citenamefont
  {Murtaza}(2014)}]{Naim+14/03}%
  \BibitemOpen
  \bibfield  {author} {\bibinfo {author} {\bibfnamefont {H.}~\bibnamefont
  {Naim}}, \bibinfo {author} {\bibfnamefont {M.~F.}\ \bibnamefont {Bashir}}, \
  and\ \bibinfo {author} {\bibfnamefont {G.}~\bibnamefont {Murtaza}},\ }\href
  {\doibase 10.1063/1.4869247} {\bibfield  {journal} {\bibinfo  {journal}
  {Phys. Plasmas}\ }\textbf {\bibinfo {volume} {21}},\ \bibinfo {eid} {032120}
  (\bibinfo {year} {2014}),\ 10.1063/1.4869247}\BibitemShut {NoStop}%
\bibitem [{\citenamefont {Bashir}\ and\ \citenamefont
  {Murtaza}(2012)}]{BashirMurtaza12/12}%
  \BibitemOpen
  \bibfield  {author} {\bibinfo {author} {\bibfnamefont {M.~F.}\ \bibnamefont
  {Bashir}}\ and\ \bibinfo {author} {\bibfnamefont {G.}~\bibnamefont
  {Murtaza}},\ }\href {\doibase 10.1007/s13538-012-0087-9} {\bibfield
  {journal} {\bibinfo  {journal} {Braz. J. Phys.}\ }\textbf {\bibinfo {volume}
  {42}},\ \bibinfo {pages} {487} (\bibinfo {year} {2012})}\BibitemShut
  {NoStop}%
\bibitem [{\citenamefont {Klein}\ and\ \citenamefont
  {Howes}(2015)}]{KleinHowes15/03}%
  \BibitemOpen
  \bibfield  {author} {\bibinfo {author} {\bibfnamefont {K.~G.}\ \bibnamefont
  {Klein}}\ and\ \bibinfo {author} {\bibfnamefont {G.~G.}\ \bibnamefont
  {Howes}},\ }\href {\doibase 10.1063/1.4914933} {\bibfield  {journal}
  {\bibinfo  {journal} {Phys. Plasmas}\ }\textbf {\bibinfo {volume} {22}},\
  \bibinfo {eid} {032903} (\bibinfo {year} {2015}),\
  10.1063/1.4914933}\BibitemShut {NoStop}%
\bibitem [{\citenamefont {Gary}(2015)}]{Gary15/04}%
  \BibitemOpen
  \bibfield  {author} {\bibinfo {author} {\bibfnamefont {S.~P.}\ \bibnamefont
  {Gary}},\ }\href {\doibase 10.1098/rsta.2014.0149} {\bibfield  {journal}
  {\bibinfo  {journal} {{Phil. Trans. R. Soc. A}}\ }\textbf {\bibinfo {volume}
  {373}} (\bibinfo {year} {2015}),\ 10.1098/rsta.2014.0149}\BibitemShut
  {NoStop}%
\bibitem [{\citenamefont {Summers}, \citenamefont {Xue},\ and\ \citenamefont
  {Thorne}(1994)}]{Summers+94/06}%
  \BibitemOpen
  \bibfield  {author} {\bibinfo {author} {\bibfnamefont {D.}~\bibnamefont
  {Summers}}, \bibinfo {author} {\bibfnamefont {S.}~\bibnamefont {Xue}}, \ and\
  \bibinfo {author} {\bibfnamefont {R.~M.}\ \bibnamefont {Thorne}},\ }\href
  {\doibase 10.1063/1.870656} {\bibfield  {journal} {\bibinfo  {journal} {Phys.
  Plasmas}\ }\textbf {\bibinfo {volume} {1}},\ \bibinfo {pages} {2012}
  (\bibinfo {year} {1994})}\BibitemShut {NoStop}%
\bibitem [{\citenamefont {Basu}(2009)}]{Basu09/05}%
  \BibitemOpen
  \bibfield  {author} {\bibinfo {author} {\bibfnamefont {B.}~\bibnamefont
  {Basu}},\ }\href {\doibase 10.1063/1.3132629} {\bibfield  {journal} {\bibinfo
   {journal} {Phys. Plasmas}\ }\textbf {\bibinfo {volume} {16}},\ \bibinfo
  {eid} {052106} (\bibinfo {year} {2009}),\ 10.1063/1.3132629}\BibitemShut
  {NoStop}%
\bibitem [{\citenamefont {Liu}\ \emph {et~al.}(2014)\citenamefont {Liu},
  \citenamefont {Liu}, \citenamefont {Dai},\ and\ \citenamefont
  {Xue}}]{Liu+14/03}%
  \BibitemOpen
  \bibfield  {author} {\bibinfo {author} {\bibfnamefont {Y.}~\bibnamefont
  {Liu}}, \bibinfo {author} {\bibfnamefont {S.~Q.}\ \bibnamefont {Liu}},
  \bibinfo {author} {\bibfnamefont {B.}~\bibnamefont {Dai}}, \ and\ \bibinfo
  {author} {\bibfnamefont {T.~L.}\ \bibnamefont {Xue}},\ }\href {\doibase
  10.1063/1.4869243} {\bibfield  {journal} {\bibinfo  {journal} {Phys.
  Plasmas}\ }\textbf {\bibinfo {volume} {21}},\ \bibinfo {eid} {032125}
  (\bibinfo {year} {2014}),\ 10.1063/1.4869243}\BibitemShut {NoStop}%
\bibitem [{\citenamefont {Astfalk}, \citenamefont {Görler},\ and\
  \citenamefont {Jenko}(2015)}]{Astfalk+15/09}%
  \BibitemOpen
  \bibfield  {author} {\bibinfo {author} {\bibfnamefont {P.}~\bibnamefont
  {Astfalk}}, \bibinfo {author} {\bibfnamefont {T.}~\bibnamefont {Görler}}, \
  and\ \bibinfo {author} {\bibfnamefont {F.}~\bibnamefont {Jenko}},\ }\href
  {\doibase 10.1002/2015JA021507} {\bibfield  {journal} {\bibinfo  {journal}
  {J. Geophys. Res.}\ }\textbf {\bibinfo {volume} {120}} (\bibinfo {year}
  {2015}),\ 10.1002/2015JA021507}\BibitemShut {NoStop}%
\bibitem [{\citenamefont {Cattaert}, \citenamefont {Hellberg},\ and\
  \citenamefont {Mace}(2007)}]{Cattaert+07/08}%
  \BibitemOpen
  \bibfield  {author} {\bibinfo {author} {\bibfnamefont {T.}~\bibnamefont
  {Cattaert}}, \bibinfo {author} {\bibfnamefont {M.~A.}\ \bibnamefont
  {Hellberg}}, \ and\ \bibinfo {author} {\bibfnamefont {R.~L.}\ \bibnamefont
  {Mace}},\ }\href {\doibase 10.1063/1.2766647} {\bibfield  {journal} {\bibinfo
   {journal} {Phys. Plasmas}\ }\textbf {\bibinfo {volume} {14}},\ \bibinfo
  {eid} {082111} (\bibinfo {year} {2007})}\BibitemShut {NoStop}%
\bibitem [{\citenamefont {Sugiyama}\ \emph {et~al.}(2015)\citenamefont
  {Sugiyama}, \citenamefont {Singh}, \citenamefont {Omura}, \citenamefont
  {Shoji}, \citenamefont {Nunn},\ and\ \citenamefont
  {Summers}}]{Sugiyama+15/10}%
  \BibitemOpen
  \bibfield  {author} {\bibinfo {author} {\bibfnamefont {H.}~\bibnamefont
  {Sugiyama}}, \bibinfo {author} {\bibfnamefont {S.}~\bibnamefont {Singh}},
  \bibinfo {author} {\bibfnamefont {Y.}~\bibnamefont {Omura}}, \bibinfo
  {author} {\bibfnamefont {M.}~\bibnamefont {Shoji}}, \bibinfo {author}
  {\bibfnamefont {D.}~\bibnamefont {Nunn}}, \ and\ \bibinfo {author}
  {\bibfnamefont {D.}~\bibnamefont {Summers}},\ }\href {\doibase
  10.1002/2015JA021346} {\bibfield  {journal} {\bibinfo  {journal} {J. Geophys.
  Res.}\ }\textbf {\bibinfo {volume} {120}} (\bibinfo {year} {2015}),\
  10.1002/2015JA021346}\BibitemShut {NoStop}%
\bibitem [{\citenamefont {Olver}\ and\ \citenamefont
  {Maximon}(2010)}]{OlverMaximon-NIST10}%
  \BibitemOpen
  \bibfield  {author} {\bibinfo {author} {\bibfnamefont {F.~W.~J.}\
  \bibnamefont {Olver}}\ and\ \bibinfo {author} {\bibfnamefont {L.~C.}\
  \bibnamefont {Maximon}},\ }in\  \cite{NIST10},\ Chap.~\bibinfo {chapter}
  {10}, pp.\ \bibinfo {pages} {215--286}\BibitemShut {NoStop}%
\bibitem [{\citenamefont {Olver}\ \emph {et~al.}(2010)\citenamefont {Olver},
  \citenamefont {Lozier}, \citenamefont {Boisvert},\ and\ \citenamefont
  {Clark}}]{NIST10}%
  \BibitemOpen
  \bibinfo {editor} {\bibfnamefont {F.~W.~J.}\ \bibnamefont {Olver}}, \bibinfo
  {editor} {\bibfnamefont {D.~W.}\ \bibnamefont {Lozier}}, \bibinfo {editor}
  {\bibfnamefont {R.~F.}\ \bibnamefont {Boisvert}}, \ and\ \bibinfo {editor}
  {\bibfnamefont {C.~W.}\ \bibnamefont {Clark}},\ eds.,\ \href
  {http://dlmf.nist.gov/} {\emph {\bibinfo {title} {{NIST Handbook of
  Mathematical Functions}}}}\ (\bibinfo  {publisher} {Cambridge},\ \bibinfo
  {address} {New York},\ \bibinfo {year} {2010})\BibitemShut {NoStop}%
\bibitem [{\citenamefont {Roy}\ \emph {et~al.}(2010)\citenamefont {Roy},
  \citenamefont {Olver}, \citenamefont {Askey},\ and\ \citenamefont
  {Wong}}]{Roy+-NIST10}%
  \BibitemOpen
  \bibfield  {author} {\bibinfo {author} {\bibfnamefont {R.}~\bibnamefont
  {Roy}}, \bibinfo {author} {\bibfnamefont {F.~W.~J.}\ \bibnamefont {Olver}},
  \bibinfo {author} {\bibfnamefont {R.~A.}\ \bibnamefont {Askey}}, \ and\
  \bibinfo {author} {\bibfnamefont {R.}~\bibnamefont {Wong}},\ }in\
  \cite{NIST10},\ Chap.~\bibinfo {chapter} {1}, pp.\ \bibinfo {pages}
  {1--39}\BibitemShut {NoStop}%
\bibitem [{\citenamefont {Gradshteyn}\ and\ \citenamefont
  {Ryzhik}(2007)}]{GradshteynRyzhik07}%
  \BibitemOpen
  \bibfield  {author} {\bibinfo {author} {\bibfnamefont {I.~S.}\ \bibnamefont
  {Gradshteyn}}\ and\ \bibinfo {author} {\bibfnamefont {I.~M.}\ \bibnamefont
  {Ryzhik}},\ }\href@noop {} {\emph {\bibinfo {title} {{Table of Integrals,
  Series and Products}}}},\ \bibinfo {edition} {seventh}\ ed.\ (\bibinfo
  {publisher} {Academic Press},\ \bibinfo {address} {San Francisco},\ \bibinfo
  {year} {2007})\ \bibinfo {note} {1222 pp.}\BibitemShut {Stop}%
\bibitem [{\citenamefont {Askey}\ and\ \citenamefont
  {Roy}(2010)}]{AskeyRoy-NIST10}%
  \BibitemOpen
  \bibfield  {author} {\bibinfo {author} {\bibfnamefont {R.~A.}\ \bibnamefont
  {Askey}}\ and\ \bibinfo {author} {\bibfnamefont {R.}~\bibnamefont {Roy}},\
  }in\  \cite{NIST10},\ Chap.~\bibinfo {chapter} {5}, pp.\ \bibinfo {pages}
  {135--147}\BibitemShut {NoStop}%
\bibitem [{\citenamefont {Wolfram}({\natexlab{a}})}]{1F2-RR-1}%
  \BibitemOpen
  \bibfield  {author} {\bibinfo {author} {\bibfnamefont {S.}~\bibnamefont
  {Wolfram}},\ }\href
  {http://functions.wolfram.com/HypergeometricFunctions/Hypergeometric1F2/17/01/01/0001/}
  {\enquote {\bibinfo {title} {{Generalized Hypergeometric Function
  $\pFq{1}{2}$}},}\ }\bibinfo {howpublished}
  {\url{http://functions.wolfram.com/HypergeometricFunctions/Hypergeometric1F2/17/01/01/0001/}}
  ({\natexlab{a}}),\ \bibinfo {note} {accessed in {\today}}\BibitemShut
  {NoStop}%
\bibitem [{\citenamefont {Daalhuis}(2010{\natexlab{a}})}]{Daalhuis-NIST10a}%
  \BibitemOpen
  \bibfield  {author} {\bibinfo {author} {\bibfnamefont {A.~B.~O.}\
  \bibnamefont {Daalhuis}},\ }in\  \cite{NIST10},\ Chap.~\bibinfo {chapter}
  {13}, pp.\ \bibinfo {pages} {321--349}\BibitemShut {NoStop}%
\bibitem [{\citenamefont {Peratt}(1984)}]{Peratt84/03}%
  \BibitemOpen
  \bibfield  {author} {\bibinfo {author} {\bibfnamefont {A.~L.}\ \bibnamefont
  {Peratt}},\ }\href {\doibase 10.1063/1.526182} {\bibfield  {journal}
  {\bibinfo  {journal} {J. Math. Phys.}\ }\textbf {\bibinfo {volume} {25}},\
  \bibinfo {pages} {466} (\bibinfo {year} {1984})}\BibitemShut {NoStop}%
\bibitem [{\citenamefont {Ablowitz}\ and\ \citenamefont
  {Fokas}(2003)}]{AblowitzFokas03}%
  \BibitemOpen
  \bibfield  {author} {\bibinfo {author} {\bibfnamefont {M.~J.}\ \bibnamefont
  {Ablowitz}}\ and\ \bibinfo {author} {\bibfnamefont {A.~S.}\ \bibnamefont
  {Fokas}},\ }\href {http://www.cambridge.org/9780521534291} {\emph {\bibinfo
  {title} {{Complex Variables. Introduction and Applications}}}},\ \bibinfo
  {edition} {2nd}\ ed.,\ {Cambridge Texts in Applied Mathematics}\ (\bibinfo
  {publisher} {Cambridge},\ \bibinfo {address} {New York},\ \bibinfo {year}
  {2003})\ \bibinfo {note} {647 pp.}\BibitemShut {Stop}%
\bibitem [{\citenamefont {Brychkov}(2008)}]{Brychkov08}%
  \BibitemOpen
  \bibfield  {author} {\bibinfo {author} {\bibfnamefont {I.}~\bibnamefont
  {Brychkov}},\ }\href {http://books.google.com.br/books?id=JQpV0bMiwSQC}
  {\emph {\bibinfo {title} {{Handbook of Special Functions: Derivatives,
  Integrals, Series and Other Formulas}}}}\ (\bibinfo  {publisher} {CRC
  Press},\ \bibinfo {year} {2008})\ \bibinfo {note} {680 + xx pp.}\BibitemShut
  {Stop}%
\bibitem [{\citenamefont {Daalhuis}(2010{\natexlab{b}})}]{Daalhuis-NIST10b}%
  \BibitemOpen
  \bibfield  {author} {\bibinfo {author} {\bibfnamefont {A.~B.~O.}\
  \bibnamefont {Daalhuis}},\ }in\  \cite{NIST10},\ Chap.~\bibinfo {chapter}
  {15}, pp.\ \bibinfo {pages} {383--401}\BibitemShut {NoStop}%
\bibitem [{\citenamefont {Wolfram}({\natexlab{b}})}]{2F1-RR-1}%
  \BibitemOpen
  \bibfield  {author} {\bibinfo {author} {\bibfnamefont {S.}~\bibnamefont
  {Wolfram}},\ }\href
  {http://functions.wolfram.com/HypergeometricFunctions/Hypergeometric2F1/17/01/01/0003/}
  {\enquote {\bibinfo {title} {{Generalized Hypergeometric Function
  $\pFq{2}{1}$}},}\ }\bibinfo {howpublished}
  {\url{http://functions.wolfram.com/HypergeometricFunctions/Hypergeometric2F1/17/01/01/0003/}}
  ({\natexlab{b}}),\ \bibinfo {note} {accessed in {\today}}\BibitemShut
  {NoStop}%
\bibitem [{\citenamefont {Hellberg}\ and\ \citenamefont
  {Mace}(2002)}]{HellbergMace02/05}%
  \BibitemOpen
  \bibfield  {author} {\bibinfo {author} {\bibfnamefont {M.~A.}\ \bibnamefont
  {Hellberg}}\ and\ \bibinfo {author} {\bibfnamefont {R.~L.}\ \bibnamefont
  {Mace}},\ }\href {\doibase 10.1063/1.1462636} {\bibfield  {journal} {\bibinfo
   {journal} {Phys. Plasmas}\ }\textbf {\bibinfo {volume} {9}},\ \bibinfo
  {pages} {1495} (\bibinfo {year} {2002})}\BibitemShut {NoStop}%
\bibitem [{\citenamefont {Luke}(1975)}]{Luke75}%
  \BibitemOpen
  \bibfield  {author} {\bibinfo {author} {\bibfnamefont {Y.~L.}\ \bibnamefont
  {Luke}},\ }\href {http://books.google.com.br/books?id=V5D6DOaCO8YC} {\emph
  {\bibinfo {title} {{Mathematical functions and their approximations}}}}\
  (\bibinfo  {publisher} {Academic Press},\ \bibinfo {address} {New York},\
  \bibinfo {year} {1975})\ \bibinfo {note} {568 + xvii pp.}\BibitemShut {Stop}%
\bibitem [{\citenamefont {Johansson}(2014)}]{mpmath}%
  \BibitemOpen
  \bibfield  {author} {\bibinfo {author} {\bibfnamefont {F.}~\bibnamefont
  {Johansson}},\ }\href {http://mpmath.org/} {\enquote {\bibinfo {title}
  {{Mpmath: a {P}ython library for arbitrary-precision floating-point
  arithmetic (version 0.19)}},}\ }\bibinfo {howpublished} {http://mpmath.org/}
  (\bibinfo {year} {2014})\BibitemShut {NoStop}%
\bibitem [{\citenamefont {Prudnikov}\ \emph {et~al.}(1990)\citenamefont
  {Prudnikov}, \citenamefont {Brychkov}, \citenamefont {Brychkov},\ and\
  \citenamefont {Mari\v{c}ev}}]{Prudnikov90v3}%
  \BibitemOpen
  \bibfield  {author} {\bibinfo {author} {\bibfnamefont {A.~A.~P.}\
  \bibnamefont {Prudnikov}}, \bibinfo {author} {\bibfnamefont {Y.~A.}\
  \bibnamefont {Brychkov}}, \bibinfo {author} {\bibfnamefont {I.~U.~A.}\
  \bibnamefont {Brychkov}}, \ and\ \bibinfo {author} {\bibfnamefont {O.~I.}\
  \bibnamefont {Mari\v{c}ev}},\ }\href
  {http://books.google.com.br/books?id=OdS6QgAACAAJ} {\emph {\bibinfo {title}
  {{Integrals and Series: More special functions}}}},\ \bibinfo {series}
  {{Integrals and Series}}, Vol.~\bibinfo {volume} {3}\ (\bibinfo  {publisher}
  {Gordon and Breach Science Publishers},\ \bibinfo {year} {1990})\ \bibinfo
  {note} {800 pp}\BibitemShut {NoStop}%
\bibitem [{\citenamefont {Agarwal}(1965)}]{Agarwal65}%
  \BibitemOpen
  \bibfield  {author} {\bibinfo {author} {\bibfnamefont {R.~P.}\ \bibnamefont
  {Agarwal}},\ }\href@noop {} {\bibfield  {journal} {\bibinfo  {journal} {Proc.
  Natl. Inst. Sci. India}\ }\textbf {\bibinfo {volume} {31}},\ \bibinfo {pages}
  {536} (\bibinfo {year} {1965})}\BibitemShut {NoStop}%
\bibitem [{\citenamefont {Sharma}(1965)}]{Sharma65b}%
  \BibitemOpen
  \bibfield  {author} {\bibinfo {author} {\bibfnamefont {B.~L.}\ \bibnamefont
  {Sharma}},\ }\href@noop {} {\bibfield  {journal} {\bibinfo  {journal} {Ann.
  Soc. Sci. Bruxelles}\ }\textbf {\bibinfo {volume} {79}},\ \bibinfo {pages}
  {26} (\bibinfo {year} {1965})}\BibitemShut {NoStop}%
\bibitem [{\citenamefont {Verma}(1966)}]{Verma66/07}%
  \BibitemOpen
  \bibfield  {author} {\bibinfo {author} {\bibfnamefont {A.}~\bibnamefont
  {Verma}},\ }\href {\doibase 10.1090/S0025-5718-66-99930-3} {\bibfield
  {journal} {\bibinfo  {journal} {Math. Comp.}\ }\textbf {\bibinfo {volume}
  {20}},\ \bibinfo {pages} {413} (\bibinfo {year} {1966})}\BibitemShut
  {NoStop}%
\bibitem [{\citenamefont {Hai}\ and\ \citenamefont
  {Yakubovich}(1992)}]{HaiYakubovich92}%
  \BibitemOpen
  \bibfield  {author} {\bibinfo {author} {\bibfnamefont {N.~T.}\ \bibnamefont
  {Hai}}\ and\ \bibinfo {author} {\bibfnamefont {S.~B.}\ \bibnamefont
  {Yakubovich}},\ }\href {http://books.google.com.br/books?id=ZC9FnwEACAAJ}
  {\emph {\bibinfo {title} {{The Double Mellin-Barnes Type Integrals and Their
  Applications to Convolution Theory}}}},\ \bibinfo {series} {{Series on Soviet
  and East European mathematics}}, Vol.~\bibinfo {volume} {6}\ (\bibinfo
  {publisher} {World Scientific},\ \bibinfo {address} {Singapore},\ \bibinfo
  {year} {1992})\ \bibinfo {note} {295 + x pp.}\BibitemShut {Stop}%
\end{thebibliography}%

\end{document}